\documentclass[aps,prx,preprint]{revtex4-2}

\usepackage{graphicx} % standard LaTeX graphics tool
\usepackage{amsmath}
\usepackage{stmaryrd}
\usepackage{MnSymbol}
\usepackage{setspace}
\usepackage{color}
\begin{document}

\title{Experimental observation of subabsorption}
\author{D. C. Gold, U. Saglam, S. Carpenter, A. Yadav, M. Beede, T. G. Walker, M. Saffman, and D. D. Yavuz}
\affiliation{Department of Physics, 1150 University Avenue,
University of Wisconsin at Madison, Madison, WI, 53706}
\date{\today}

\begin{abstract}

We predict and experimentally demonstrate a new type of collective (cooperative) coupling effect where a disordered atomic ensemble absorbs light with a rise-time longer (i. e., at a rate slower) than what is dictated by single-atom physics. This effect, which we name subabsorption, can be viewed as the absorptive analog of subradiance. The experiment is performed using a dilute ensemble of ultracold $^{87}$Rb atoms with a low optical depth, and time-resolving the absorption of a weak (tens of photons per pulse) resonant laser beam.   In this dilute regime, the collective interaction relies on establishing dipole-dipole correlations over many atoms; i.e., the interaction is not dominated by the nearest neighbors. As a result,  subabsorption is highly susceptible to motional dephasing: even a temperature increase of 60 $\mu$K is enough to completely extinguish the subabsorption signal. We also present a theoretical model whose results are in reasonable agreement with the experimental observations. The model uses density-dependent dephasing rate of the long-range dipole-dipole correlations as a single adjustable parameter. Experiment-theory comparison indicates a dephasing coefficient of $\beta/2 \pi  = 4.9 \times 10^{-5}$ Hz~cm$^3$, which is more than two orders of magnitude larger than the known dipole-dipole line broadening coefficient in $^{87}$Rb.   

\end{abstract} 
\maketitle
\newpage

\section{Introduction and background}

Interaction of light with atoms plays an important role in a wide range of physical processes that we observe in nature. Since the pioneering work of Dicke over 70 years ago, we know that an ensemble of atoms may interact with light in a way that is qualitatively different than how a single atom interacts \cite{dicke,haroche}. This is due to the collective (cooperative) coupling of the atoms to light and studies of such collective interaction is still an active research area \cite{scully1,yelin,francis,adams,jenkins,ritsch,bachelard,petrov,zanthier,agarwal,reitz,bennett,cirac,pohl}. If we have many atoms per cubic wavelength of the emitted radiation, the ensemble can spontaneously decay at a rate that is much faster than the single-atom decay rate. This is typically referred to as Dicke superradiance \cite{feld,manassah}, and has been experimentally observed in many different physical systems, including neutral atoms \cite{bloch,an,gauthier,kuga}, ions \cite{ions}, molecules \cite{molecules}, and various solid state systems \cite{diamond1,diamond2,superconducting,quantumdot}. Superradiance relies on constructive interference of the emitted radiation from individual atoms. It is also well-known that, under certain conditions, the emissions can instead interfere destructively, causing reduced decay rates. This is referred to as subradiance and has largely been studied using laser-cooled ultracold ensembles \cite{kaiser1,kaiser2,kaiser3,kaiser4,kaiser5,browaeys1,browaeys2,browaeys3}. 

Studies of collective coupling have historically focused on cooperative effects that happen when the atoms start in their excited state; i.e., superradiance and subradiance. These studies have a long history, with early experimental studies of superradiance in atomic vapors dating back to the 1970s \cite{feld,manassah}.  While much less known, it is expected that these effects may have absorptive counterparts when instead all the atoms start in their ground state. These absorptive collective physical processes are schematically described in Fig.~1. In a recent beautiful paper, An and colleagues have predicted and experimentally demonstrated superabsorption, which can be viewed as the absorptive analog of superradiance \cite{superabsorption}. By controlling the atomic positions in a cavity using a nano-patterned mask, they demonstrated that the atomic ensemble can absorb photons in the cavity mode at a rate faster than what is dictated by single-atom physics. In this paper, we predict and experimentally demonstrate subabsorption, which can be viewed as the absorptive analog of subradiance. We start with a laser cooled ensemble of $^{87}$Rb atoms that are trapped in a magneto-optical trap. With all the atoms starting in their ground state, we shine a very weak (tens of photons per pulse), resonant laser beam on the atoms. The incident laser pulse is turned-on sharply and we measure the amount of absorbed light in the cloud as a function of time by recording the output pulse and comparing it with the incident beam. The absorption of the laser pulse through the atomic cloud has an associated rise-time, since the atoms transition to the excited level and establish a dipole moment. Without any collective interaction, this rise-time is exactly determined by the lifetime of the excited level, and is $2 \tau_a$ ($\tau_a = 1/\Gamma_a=26.2$~ns is the excited state lifetime in $^{87}$Rb). We show that this rise-time can increase by as much as 50~\% when the optical depth of the ensemble is about OD~$\sim 0.1$ and also when the ensemble is dilute (with a density of about 0.01~atoms per wavelength cube of volume). 

\begin{figure}[tbh]
\vspace{0cm}
\begin{center}
%\hspace{3cm}
\includegraphics[width=0.9\textwidth]{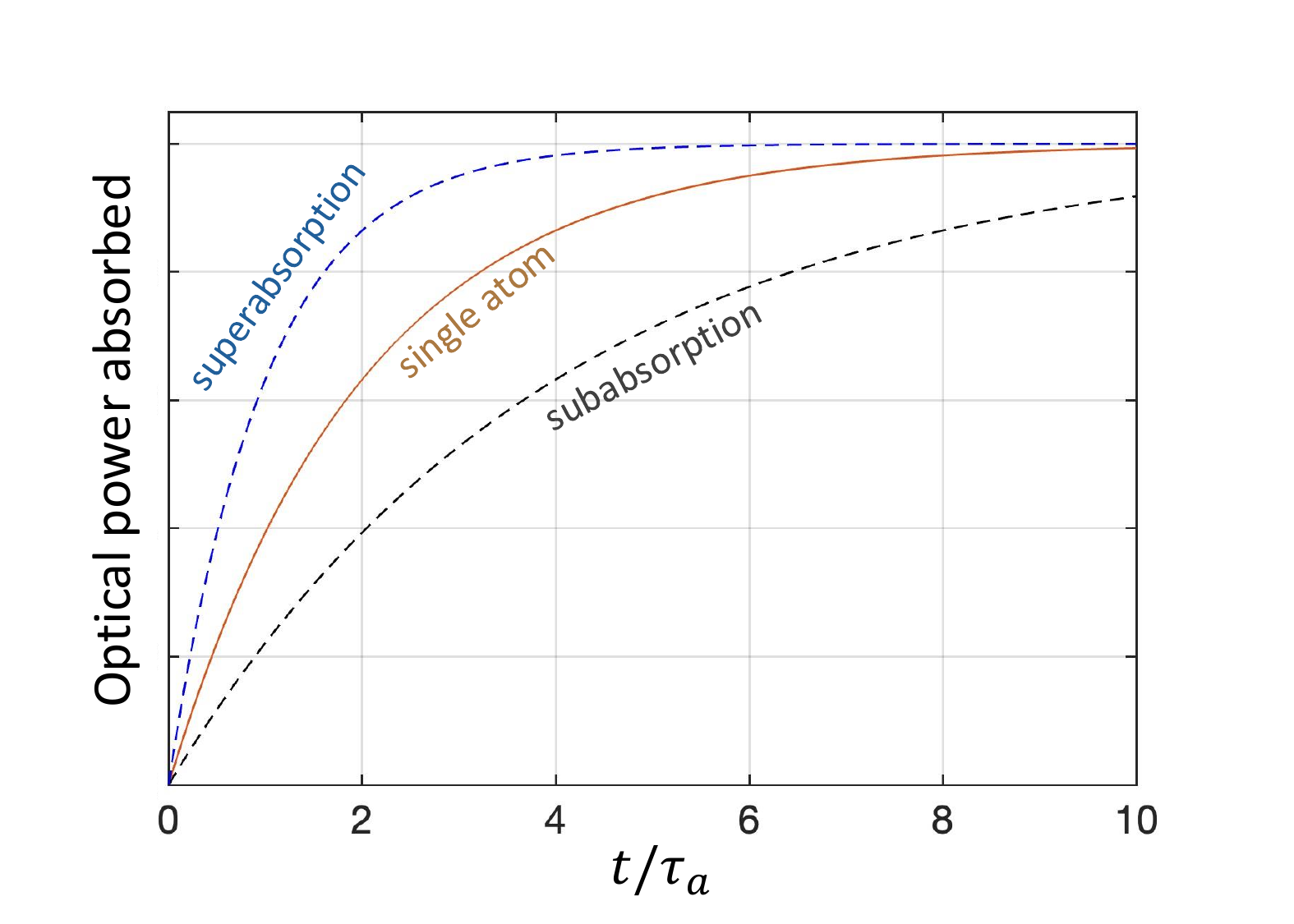}
\vspace{-0.2cm} 
\begin{singlespace}
\caption{\label{scheme} \small Qualitative description of superabsorption and subabsorption. With all the atoms starting in their ground state, when a resonant laser pulse is turned on sharply at $t=0$, the absorption does not instantaneously jump to its steady-state value.  Instead, the absorption has an associated rise-time, since the atoms transition to the excited level and establish a dipole moment. Without any collective interaction, this rise-time is exactly determined by the lifetime of the excited level, and is $2 \tau_a$ (solid curve). With collective interaction, the rise-time can be faster (superabsorption, blue dashed curve) or slower (subabsorption, black dashed curve). Our experiments demonstrate subabsorption in an ultracold disordered ensemble at a low optical depth and density. }
\end{singlespace}
\end{center}
\vspace{-0.cm}
\end{figure}

To observe the subabsorption signal, it is critical that the ensemble is dilute with a low optical depth. As we discuss below, as we increase the optical depth of the ensemble to OD$\sim 1$, the subabsorption signal vanishes. This is because, subabsorption relies on establishing atomic states that are correlated across the ensemble; i.e., it relies on establishing long range correlations. As the density and the optical depth of the ensemble increases, dipole-dipole dephasing decreases the collective coupling. We justify this viewpoint by including a single density-dependent dephasing rate in our model of the form $\gamma_{DD}= \beta n$ to explain the observed results. As we will discuss below, with a single adjustable parameter, there is reasonable agreement with the experimental results and the numerical predictions. The best experiment-theory agreement is obtained for a dephasing coefficient of $\beta/2 \pi  = 4.9 \times 10^{-5}$ Hz~cm$^3$, which is more than two orders of magnitude larger than the known dipole-dipole line broadening coefficient in $^{87}$Rb. The fact that long-range correlations are critical for the observed results is also justified by the fragility of the subabsorption signal to motional dephasing. Even a temperature increase of 60~$\mu$K of the atomic ensemble, which corresponds to a motional dephasing of $\sim \lambda_a/100$ in an atomic lifetime, is enough to largely extinguish the subabsorption signal. 

From these observations, we conclude that the physics of subabsorption is similar to subradiance \cite{dipto} and the spatial coherence of light that we recently experimentally observed \cite{davidexp} and numerically studied \cite{yavuz_coherence}, again in dilute ensembles with a low optical depth. In our recent experimental work, we had also observed that the subradiance signal peaked at a certain optical depth: further increase in the optical depth had resulted in a decrease of the subradiance signal \cite{davidexp}. Furthermore, the established long range dipole-dipole correlations resulted in the transverse spatial coherence of the light, which we characterized using a misaligned Michelson interferometer. The observed spatial coherence was quite sensitive to motional dephasing. We found that, similar to subabsorption measurements that we will discuss below, a temperature increase of $\sim 100$~$\mu$K (i.e., motional dephasing of $\sim \lambda_a/100$ in an atomic lifetime), was enough to diminish the long range correlations and the observed spatial coherence.

The outline for this paper is as follows: in Sections~II and III below, we will discuss the experimental set-up and experimental results including our measurements of absorption rise-time as a function of the optical depth and atomic temperature. In Section~IV, we will discuss our numerical simulations where we treat the atomic ensemble as a non-interacting gas (i.e., ignoring the dipole-dipole interactions) using a Maxwell-Bloch propagation code. In Section~V we will summarize the results of numerical calculations that include the long-range dipole-dipole interactions. Much of the details regarding all of the Sections will be discussed in the appendices. 

Before we proceed further, we would like to cite other pertinent prior work. There is a recent surge of interest in the studies of collective coupling, in particular due to various applications in quantum information science, such as proposals for highly directional mapping of quantum information between atoms and light in two-dimensional arrays \cite{yelin1,yelin2}, and improving photon storage fidelities using subradiance \cite{kimble}. Other highlights of recent theory work include studies of broadening and photon-induced atom recoil in collective emission \cite{francis1,francis2,francis3}, light storage in optical lattices \cite{ritsch2,ballantine,garcia},  collective nonclassical light emission  and hyperradiance \cite{agarwal2,agarwal3,agarwal4}, and superradiance in multivel systems \cite{masson}. On the experimental front, as we mentioned above, much early work on subradiance used disordered ultracold atomic clouds \cite{kaiser1,kaiser2,kaiser3,kaiser4,kaiser5,browaeys1,browaeys2,browaeys3}, including our recent work \cite{dipto,davidexp}, which used dilute
ensembles with low optical depth. Recent experiments
using ultracold atoms demonstrated single-atomic-layer mirrors \cite{bloch}, phase transitions \cite{browaeys4}, enhanced collective coupling using optical cavities \cite{yan}, and interfacing of atomic clouds with nanophotonic circuits \cite{hung} and chiral nanofibers \cite{liedl}.

\section{Experimental schematic}

Figure~2 shows a simplified schematic of our experiment. The experiment starts with a magneto-optical trap (MOT) of laser-cooled ultracold rubidium ($^{87}$Rb) atoms, which is loaded from a background vapor inside an ultrahigh vacuum chamber. Laser cooling is implemented on the $F=2 \rightarrow F'=3$ cycling transition of the D2 line of $^{87}$Rb, near a wavelength of 780 nm. Further details on the experimental set-up can be found in Appendix~A.  At the end of the MOT loading cycle, we typically trap $\sim 10^5$ atoms, within a radius of 0.25~mm. The atomic temperature is about $40$~$\mu$K, which is measured by monitoring the free expansion of the cloud using an electron-multiplying CCD camera. During the final 10 ms of the MOT loading cycle, we turn-off the hyperfine repumper beam. As a result, the atoms are optically pumped into the $F=2$ ground hyperfine level at the end of the cycle.

\begin{figure}[tbh]
\vspace{0cm}
\begin{center}
%\hspace{3cm}
\includegraphics[width=1\textwidth]{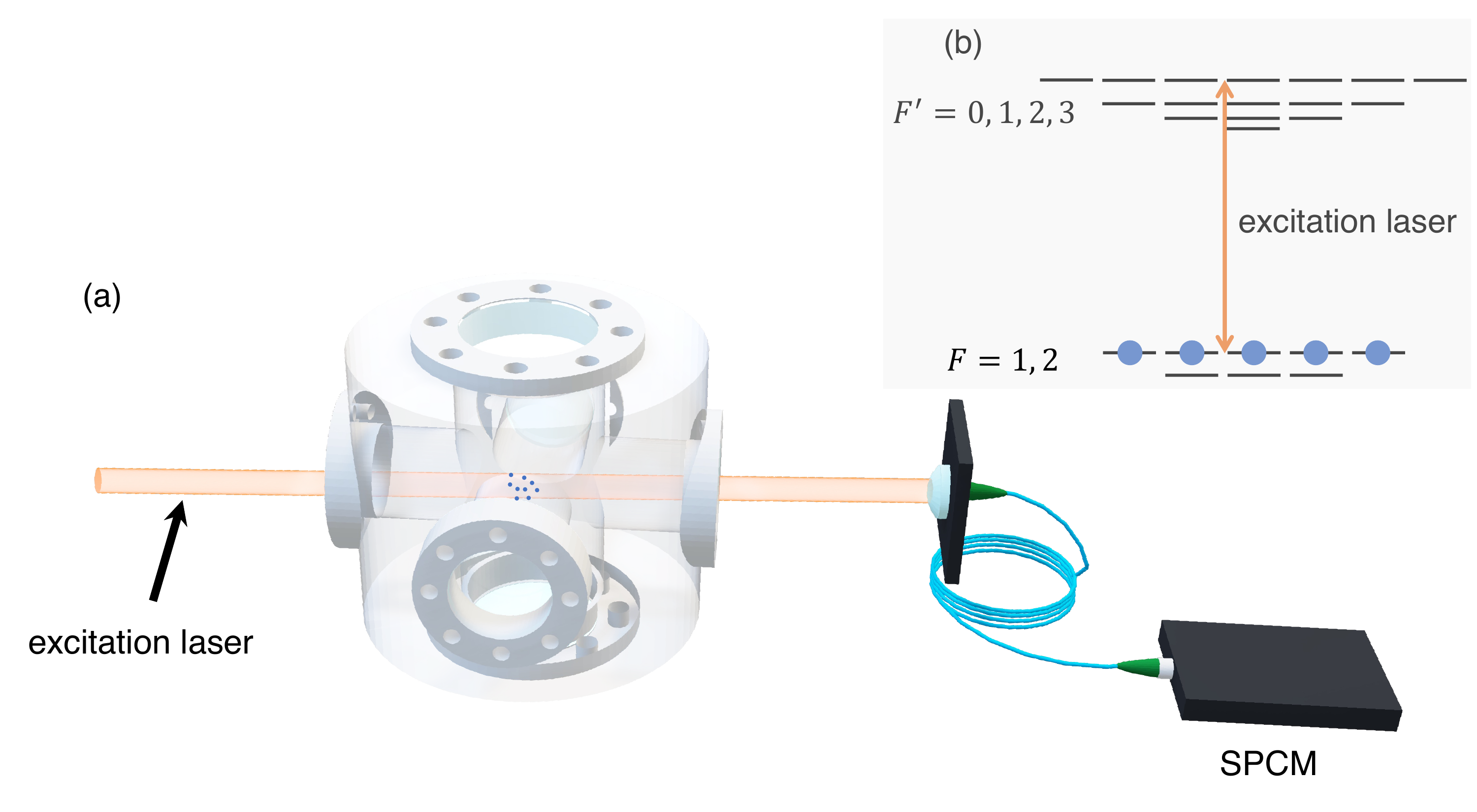}
\vspace{-0.2cm} 
\begin{singlespace}
\caption{\label{scheme} \small (a) Simplified schematic of the experiment. The experiment starts with a magneto-optical trap (MOT) of laser-cooled ultracold rubidium ($^{87}$Rb) atoms. With the atoms trapped and optically pumped to the $F=2$ ground level, we turn on a weak excitation laser which is tuned to the $F=2 \rightarrow F'=3$ cycling transition. After passing through the atomic ensemble, the excitation laser pulse is detected using a single-photon counting module (SPCM). The measurement is performed by detecting the excitation laser pulse on the photon counter with and without the atomic ensemble, and thereby time-resolving the amount of absorbed light in the cloud. (b) The relevant energy level diagram of $^{87}$Rb. The excitation laser is near a wavelength of 780~nm and is tuned to the cycling transition between the  $5 S_{1/2}$ to $5P_{3/2}$ electronic states (D2 line). }
\end{singlespace}
\end{center}
\vspace{-0.cm}
\end{figure}

With the atoms trapped and optically pumped to the $F=2$ ground level, we turn on a weak resonant laser which is tuned to the $F=2 \rightarrow F'=3$ cycling transition. This laser, termed the excitation laser, is loosely focused on the atomic cloud with a long focal length lens, and has a spatial size that is comparable to the size of the ensemble. The abrupt turning-on of the excitation laser is achieved using a fast acousto-optic modulator (AOM) with a 90~\%$-$10~\% turn-on time of 8 ns, which is significantly shorter than the excited-state lifetime of $\tau_a =26.2$~ns.  After passing through the atomic ensemble, the excitation laser pulse is detected using a single-photon counting module (SPCM). To ensure that saturation and optical pumping effects do not play a role in our experiment, we adjust the excitation laser light intensity such that the number of photons in the excitation laser pulse is at the level of tens of photons, which is much smaller than the number of atoms in the ensemble. As a result, the experiment is deeply in the weak excitation regime, with the Rabi frequency of the laser pulse much smaller than the excited state decay rate, $\Omega << \Gamma_a$. We have experimentally verified that the experiment is in the weak-excitation regime; when the intensity of the incident laser pulse is increased by an order of magnitude, the experimental results are qualitatively unchanged. 

The measurement is performed by detecting the excitation laser pulse on the photon counter with and without the atomic ensemble. The main difficulty in the experiment is to obtain enough statistics so that the absorption as a function of time curve can be measured with a good signal-to-noise ratio. This becomes challenging when the optical depth of the ensemble is very low, since the excitation laser intensity is only attenuated by, for example, a few percent. To obtain a sufficiently good signal-to-noise, a single rise-time measurement typically requires averaging over $\sim 10^5$ experimental cycles. 

We refer to the detected signal on the photon counter without the atoms as $I_{input}$ and with the atoms as $I_{output}$. We define the quantity $\sigma (t)$, which is the absorption of the excitation laser as a function of time, through the following relationship:
\begin{eqnarray}
I_{output}(t) = I_{input}(t) \exp{\left[ -\sigma(t) \right] } \quad . 
\end{eqnarray}

\noindent We note that $\sigma(t)$ can be thought as the optical depth of the ensemble as a function of time. In the experiment, by recording the input and the output pulses on the SPCM, we measure $\sigma (t)$ and estimate its rise-time by fitting it to an exponential function. Throughout the paper, we will refer to the measured rise-time with the quantity $\tau$. As we will discuss below in detail in Section~IV, single-atom physics predict $\sigma (t)$ to have the following form:
\begin{eqnarray}
\sigma(t) = \sigma_{ss} \{ 1- \exp{\left[ -t/(2 \tau_a) \right] } \} \quad . 
\end{eqnarray}

\noindent Here, the quantity $\sigma_{ss}$ is the steady-state value of the optical depth. Noting Eq.~(2), single-atom physics predicts the exponential rise-time to be exactly $\tau= 2 \tau_a = 52.4$~ns (ignoring the propagation effects that come into play at higher optical depths). In the experimental results that we discuss below, we measure this rise-time as the parameters of the ensemble such as the optical depth and the atomic temperature are varied.

\section{Experimental results}

Figure~3 shows a sample dataset where we measure the absorption as a function of time, $\sigma(t)$, for an atomic cloud with a steady-state optical depth of $\sigma_{ss} =0.87$. The trigger for the excitation laser is turned on at $t=0$, after which the excitation laser rises to its steady state value in about 8~ns. We record data points until a time of $8 \tau_a$ and fit the data to an exponential rise between $\tau_a< t < 8 \tau_a$. Further details regarding our data taking and curve fitting procedure can be found in Appendix~A.  There are two main reasons why we do not include the early time  data, $0< t < \tau_a $, in our fits and data analysis: (1) In this region, the signal-to-noise ratio is considerably worse, especially when the optical depth of the ensemble is very low. This reduction in the signal-to-noise affects the quality of the fits. (2) In some of the datasets at low optical depths, we see signatures of superabsorption in the early times followed by subabsorption: i.e., these datasets display signatures of superabsorption-to-subabsorption transition. This is similar to subradiance in dilute clouds being accompanied by early time superradiance and superradiance-to-subradiance transition \cite{dipto,davidexp,rudhy}. While superabsorption-to-subabsorption transition is very interesting, we  focus on subabsorption in the current work. We leave experimental studies of early time superabsorption as well as superabsorption-to-subabsorption transition for future work.

\begin{figure}[tbh]
\vspace{0cm}
\begin{center}
%\hspace{3cm}
\includegraphics[width=1\textwidth]{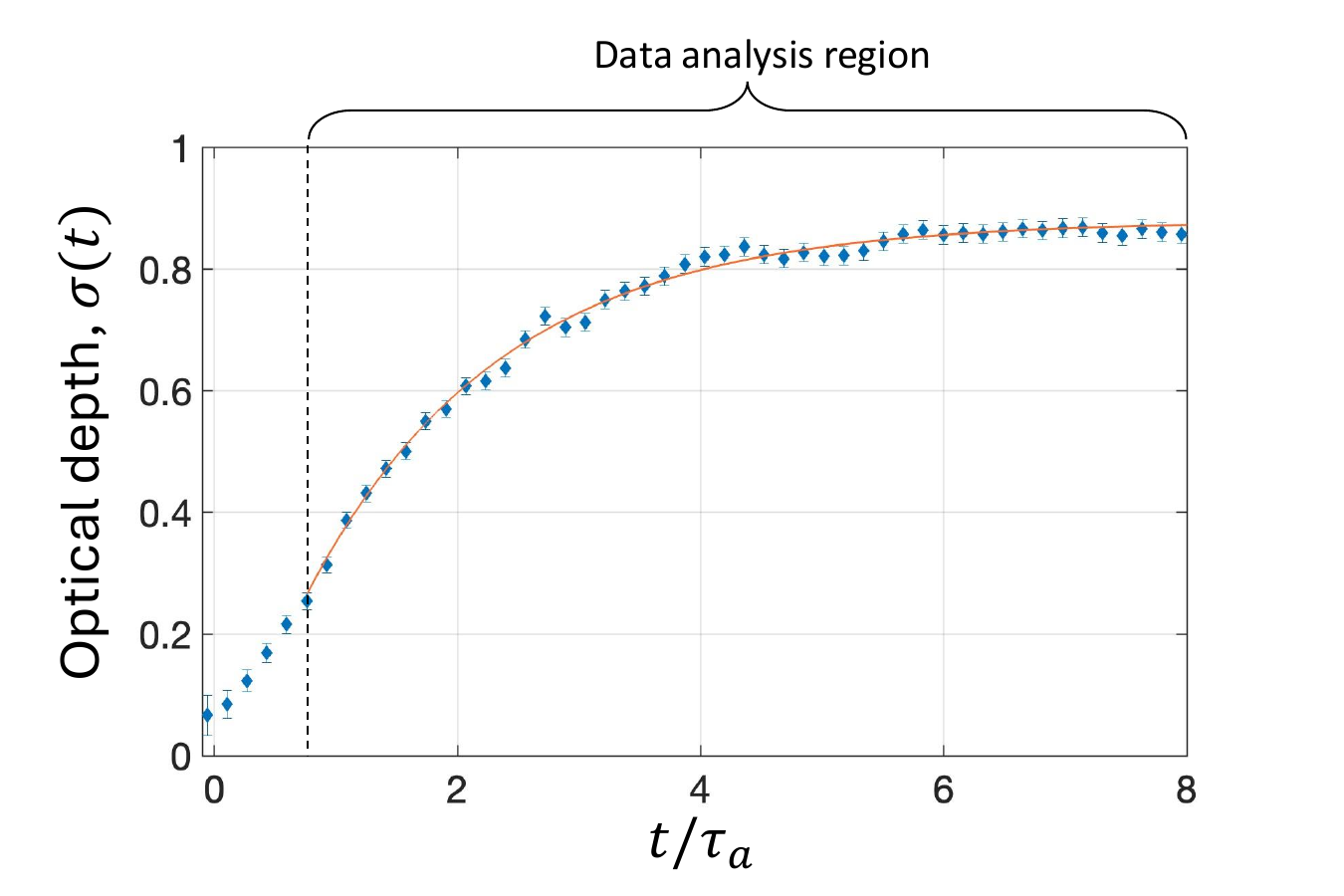}
\vspace{-0.5cm} 
\begin{singlespace}
\caption{\label{scheme} \small A sample dataset where we measure absorption as a function of time, $\sigma(t)$, for an atomic cloud with a steady-state optical depth of $\sigma_{ss} =0.87$. The trigger for the excitation laser is turned on at $t=0$, after which the excitation laser rises to its steady state value in about 8~ns. We fit the data to an exponential rise between $\tau_a< t < 8 \tau_a$. The fit for this specific dataset is shown in solid red curve. We refer to the rise-time of the exponential fit as the absorption rise-time and we denote this quantity with $\tau$. See text for further details.  }
\end{singlespace}
\end{center}
\vspace{-0.cm}
\end{figure}

We next discuss our results where we study the absorption rise-time as we vary the optical depth of the ensemble. We vary the optical depth by turning off the trapping beams for a certain amount of time and letting the cloud freely expand, thereby increasing the size of the atomic ensemble (and reducing its density and optical depth). Figure~4(a) shows the experimentally measured absorption rise-time, $\tau$ (normalized to the expected rise-time from single-atom physics of $2 \tau_a$) as the steady-state optical depth of the ensemble is varied from $\sigma_{ss} = 0.024 $ to $\sigma_{ss} = 1.11 $. The error bars on each data point are a measure of the quality of the exponential fit (further details on how the error bars are calculated can be found in Appendix~A). The red solid line in Figure~4(a) is a numerical calculation using the Maxwell-Bloch propagation code that ignores dipole-dipole correlations and models the atomic ensemble as a non-interacting gas.  We will discuss the Maxwell-Bloch propagation model in detail in Section~IV below. At very low optical depths, the Maxwell-Bloch calculation converges to $\tau= 2 \tau_a$, which is what is expected from single-atom physics. As the optical depth increases the calculated rise-time with the propagation code decreases due to propagation effects inside an optically-thick cloud becoming relevant.

 \begin{figure}[p]
\vspace{0cm}
\begin{center}
%\hspace{3cm}
\includegraphics[width=0.9\textwidth]{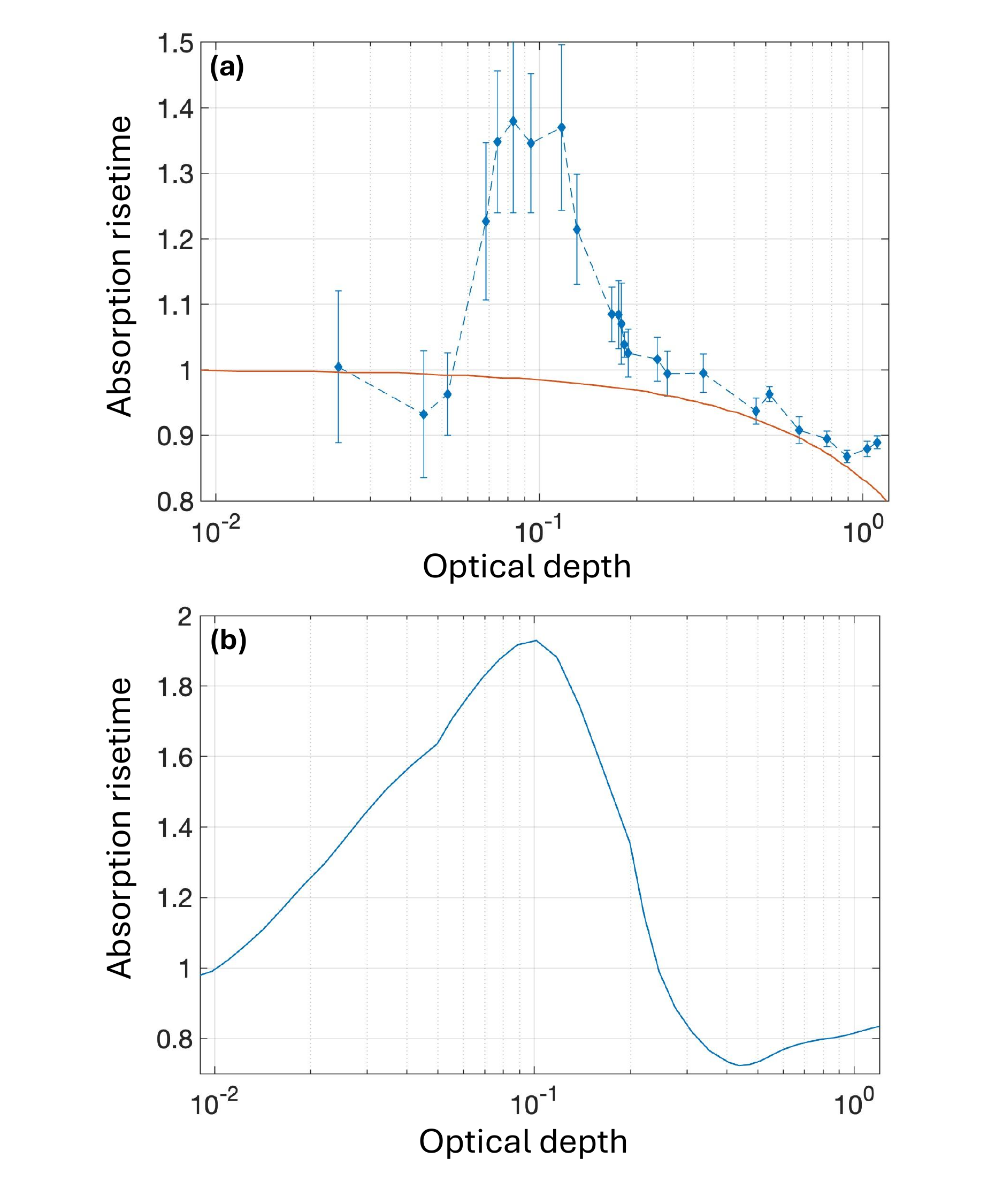}
\vspace{-0.5cm} 
\begin{singlespace}
\caption{\label{scheme} \small (a) Experimentally measured absorption rise-time, $\tau$ (normalized to $2 \tau_a$), as the steady-state optical depth of the ensemble is varied from $\sigma_{ss} = 0.024 $ to $\sigma_{ss} = 1.11 $. The solid red curve is a numerical calculation that treats the atomic ensemble as a non-interacting gas (Maxwell-Bloch propagation code). The experimental measurements agree well with the non-interacting gas calculation at low optical depths, $\sigma_{ss} < 0.05 $, and at optical depths  $\sigma_{ss} > 0.5 $. In the intermediate region, absorption rise-time $\tau$ is increased considerably near $\sigma_{ss} \sim 0.1$, showing clear evidence of subabsorption. (b) Numerical calculation including dipole-dipole correlations using density-dependent dephasing as a single adjustable fitting parameter. See text for details.  }
\end{singlespace}
\end{center}
\vspace{-0.cm}
\end{figure}

\begin{figure}[p]
\vspace{-1cm}
\begin{center}
%\hspace{3cm}
\includegraphics[width=0.9\textwidth]{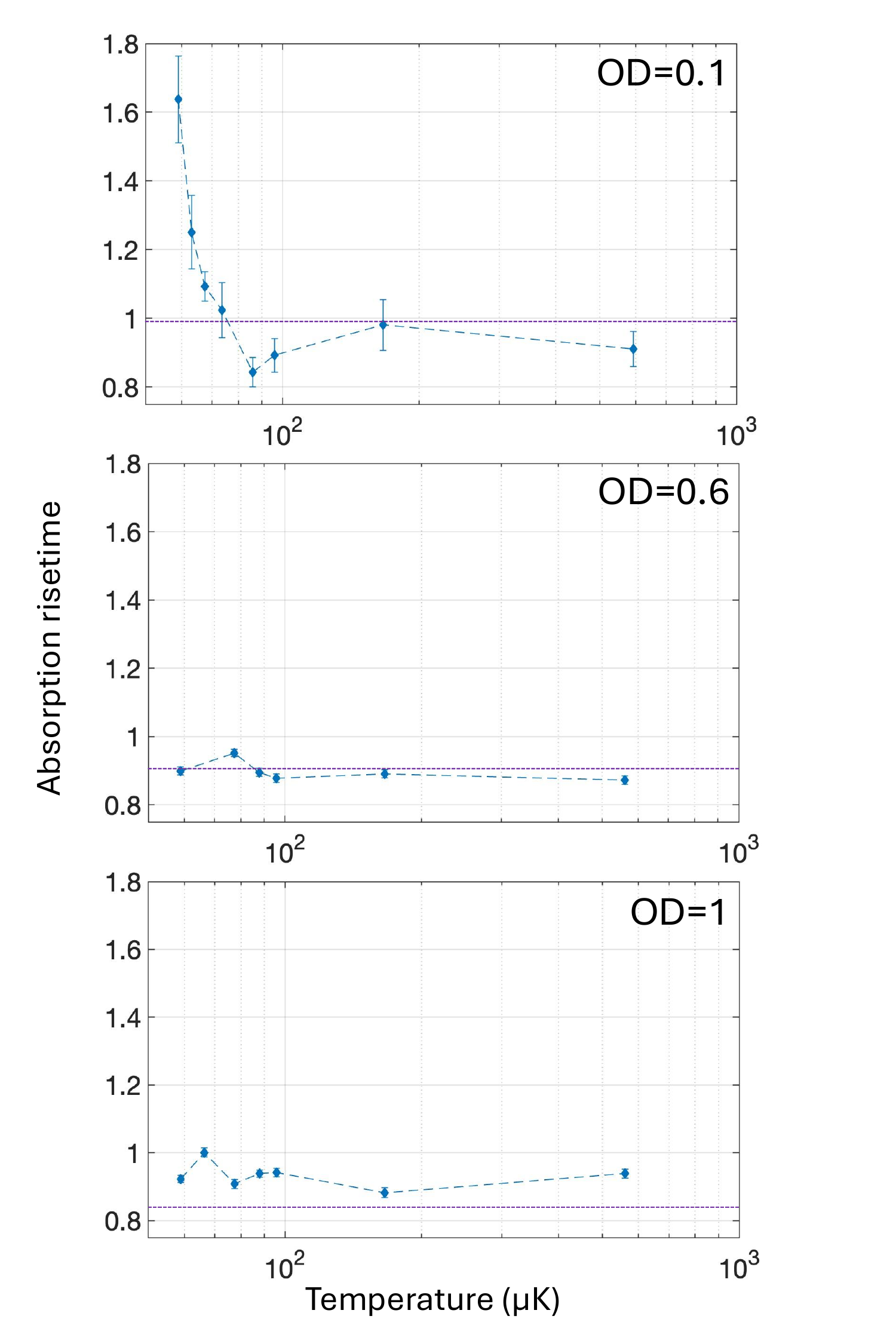}
\vspace{-0.7cm} 
\begin{singlespace}
\caption{\label{scheme} \small Experimentally observed absorption rise-time as a function of the temperature of the atomic ensemble at three different optical depths, $\sigma_{ss}=0.1$, $\sigma_{ss}=0.6$, and $\sigma_{ss}=1$, respectively. The subabsorption signal at $\sigma_{ss}=0.1$ is highly susceptible to motional dephasing and vanishes when the atomic temperature is increased to about 100~$\mu$K (corresponding to a motional dephasing of $\sim \lambda_a/100$ in an atomic lifetime). At higher optical depths, $\sigma_{ss}=0.6$, and $\sigma_{ss}=1$, there is not significant variation of the rise-time as the temperature is increased. In each plot, the horizontal dashed line is the expected rise-time from the Maxwell-Bloch propagation code.   }
\end{singlespace}
\end{center}
\vspace{-0.cm}
\end{figure}
 
Noting Fig.~4(a), the experimental measurements agree well with the non-interacting gas calculation at very low optical depths, $\sigma_{ss} < 0.05 $, and at optical depths  $\sigma_{ss} > 0.5 $. But in the intermediate region, there is large deviation from the non-interacting gas calculation. In particular, the rise-time $\tau$ is increased considerably near $\sigma_{ss} \sim 0.1$, clearly demonstrating subabsorption. Figure~4(b) is a numerical calculation that includes the dipole-dipole interactions between the atoms. This calculation uses a single adjustable parameter, which is the density-dependent dephasing of the dipole-dipole correlations, of the form $\gamma_{DD}= \beta n$ (the quantity $n$ is the density of the ensemble). With this single adjustable parameter, there is reasonable agreement between the theoretical calculation and the experimental results. This agreement is obtained for a dephasing coefficient of $\beta/2 \pi  = 4.9 \times 10^{-5}$ Hz~cm$^3$. We will discuss our numerical model that includes the dipole-dipole correlations in detail in Section~V below.

We next discuss the temperature dependence of the absorption rise-time and the fragile nature of the subabsorption signal. If the long-range dipole-dipole correlations are responsible for the observed subabsorption, we would expect the subabsorption signal to be sensitive to motional dephasing and, therefore, to the atomic temperature. Figure~5 shows the experimentally observed absorption rise-time (again normalized to $2 \tau_a$) as a function of the temperature of the atomic ensemble at three different optical depths, $\sigma_{ss}=0.1$, $\sigma_{ss}=0.6$, and $\sigma_{ss}=1$, respectively. We increase the temperature of the atomic ensemble by exposing the atoms to an intense counter-propagating beam pair that is resonant to the cycling transition, for a short duration of time (1~$\mu$s to 10~$\mu$s). The duration of the resonant pulse controls the amount of heating and therefore the atomic temperature.

For an optical depth of $\sigma_{ss}=0.1$ (the top plot in Fig.~5), we observe that the subabsorption signal disappears almost completely as the temperature of the ensemble is increased to about 100~$\mu$K. This is a clear indication that the subabsorption signal is highly susceptible to motional dephasing and suggests that many-atom correlations are responsible for the observed subabsorption. As we mentioned above, this is similar to our recent measurements where we observed that the spatial coherence in collective spontaneous emission also had  very sensitive temperature dependence \cite{davidexp}. We showed that long-range dipole-dipole correlations during subradiant decay resulted in transverse spatial coherence, which was greatly reduced when the temperature of the ensemble was increased to about 100~$\mu$K. 

For comparison, Figure~5 also shows the absorption rise-time at higher optical depths of the atomic ensemble, $\sigma_{ss}=0.6$, and $\sigma_{ss}=1$, respectively. At these higher optical depths, we do not observe significant variation of the rise-time as the temperature of the ensemble is increased. In each plot, the horizontal dashed line is the expected rise-time from the Maxwell-Bloch propagation code (i.e., modeling the ensemble as a non-interacting gas). At these higher optical depths of $\sigma_{ss}=0.6$, and $\sigma_{ss}=1$, the measurements are in reasonable agreement with what is expected from the Maxwell-Bloch code.  

\section{Non-interacting gas: Maxwell-Bloch formalism}

To model the absorption of a laser pulse through an atomic ensemble without the dipole-dipole correlations, we use the well-known coupled Maxwell-Bloch equations \cite{eberly}. We assume plane-wave excitation and consider an ensemble of two-level atoms, with states $|0 \rangle$, and $|1\rangle$. For the atomic system, the evolution of the density matrix elements in the presence of the laser pulse is:
\begin{eqnarray}
\frac{\partial \rho_{00}}{dt'} & = & \Gamma_a \rho_{11} + \frac{i}{2} \left( \Omega \rho_{01}^* - \Omega^* \rho_{01} \right) \quad , \nonumber \\ 
\frac{\partial  \rho_{11}}{dt'} & = & - \Gamma_a \rho_{11} + \frac{i}{2} \left( \Omega^* \rho_{01} - \Omega \rho_{01}^* \right) \quad , \nonumber \\
\frac{\partial \rho_{01}}{dt'} & = & -\left( \frac{\Gamma_a}{2} + \Delta \right) \rho_{01} + \frac{i}{2} \Omega \left( \rho_{11} - \rho_{00} \right)  \quad .   
\end{eqnarray}

\noindent Here, $t'$ is the local time, $t' = t- z/c$, and the quantities $\rho_{00}$ and $\rho_{11}$ are the diagonal elements of the density matrix. $\rho_{01}$ is the off-diagonal density matrix element (the coherence), $\Omega$ is the Rabi-frequency of the applied laser pulse, and $\Delta= \omega_a - \omega_{laser}$ is the detuning of the excitation laser from the two level atomic transition. With the density-matrix elements evolving according to Eq.~(3), the slowly-varying envelope propagation equation (SVEA), that dictates the propagation of the excitation laser, is:
\begin{eqnarray}
\frac{\partial \Omega}{\partial z} = - \frac{i}{\hbar} \eta \omega_{laser} n \mu_{dipole}^2 \rho_{01} \quad .  
\end{eqnarray}

\noindent Here, $\eta=\sqrt{\mu_0 / \epsilon_0}$ is the impedance of free space, $\mu_{dipole}$ is the dipole matrix element of the two-level atomic transition, and $n$ is the atomic density. In our simulations, we solve the above partial differential equations for $\rho_{00} (z,t')$, $\rho_{11} (z,t')$, $\rho_{01} (z,t')$, and $\Omega (z,t')$ numerically in a space-time grid.  In the numerical results that are displayed in the solid red line of Fig.~4(a), we solve these equations with the initial condition that all the atoms in the ensemble start in the ground state, $\rho_{00} (z,t'=0)=1$, $\rho_{11} (z,t'=0)=0$, $\rho_{01} (z,t'=0)=0$, and the boundary condition for the applied excitation laser $\Omega (z=0,t')= \Omega_{input}(t')$. Further details regarding these numerical simulations are given in Appendix~B.

The analytical solution that is displayed in Eqs.~(1) and (2) can be derived from the above set of coupled differential equations in the perturbative (weak-excitation) limit. When the Rabi-frequency of the excitation laser is much smaller than the atomic decay rate, $\vert \Omega \vert << \Gamma_a$, we can assume $\rho_{00} \approx 1$ and $\rho_{11} \approx 0$. In this limit, the solution for the off-diagonal density matrix element for on-resonance excitation ($\Delta=0$) is:
\begin{eqnarray}
    \rho_{01} =- i  \frac{\Omega}{\Gamma_a} \{ 1 -\exp{ \left[ -(\Gamma_a/2) t' \right]} \} \quad . 
\end{eqnarray}

\noindent With the solution for the off-diagonal density matrix element known, the solution for the excitation laser pulse [Eq.~(4)], as a function of propagation distance, $z$, is:
\begin{eqnarray}
    \Omega(z, t') = \Omega(z=0, t') \exp{ \left( - \frac{\alpha}{2} \{ 1 -\exp{ \left[ -(\Gamma_a/2) t' \right]} \} z \right)  }  \quad ,  
\end{eqnarray}

\noindent where we have defined the absorption constant, $\alpha = \eta \omega_{laser} n \mu_{dipole}^2/ \hbar \Gamma_a $. For a medium of length $L$, we obtain Eqs.~(1) and (2) by evaluating the above expression at $z=L$ and relating the input and output intensity to the Rabi frequency by using: $I_{input} (t')= \vert \Omega(z=0, t') \vert^2 \hbar^2/ 2 \eta \mu_{dipole}^2 $ and $I_{output} (t')= \vert \Omega(z=L, t') \vert^2 \hbar^2/ 2 \eta \mu_{dipole}^2  $. We note that, in our experiment, the spatial size of the atomic ensemble is small, less than 1~mm. As a result, the difference between time and local time is not of importance for the time scales of our experiment and we can replace $I_{input}(t) = I_{input}(t')$ and $I_{output}(t) = I_{output}(t')$.

In the above formulation, we have assumed the frequency of the excitation laser to be exactly resonant with the two-level atomic transition, $\Delta = 0$. For nonzero detuning, $\Delta \neq 0$, the absorption curves deviate substantially from a simple exponential rise. We have studied the effect of nonzero detuning both experimentally and numerically, which we discuss in detail in Appendix~D below. Even a frequency detuning of $\Delta = \Gamma_a/3$ can decrease the absorption rise-time by 35\%. 

\section{Numerical Simulations with Dipole-Dipole Coupling}

In this Section, we describe our numerical results where we model the absorption of the cloud while taking into account long-range dipole-dipole interactions. In our model, there are two key differences compared to more-traditional numerical simulations of collective decay from atomic ensembles \cite{yavuz_coherence}: (1) The ensemble starts with all the atoms in their ground level (instead of the excited level), and (2) in addition to the dipole-dipole interactions between the atoms that are mediated by vacuum electromagnetic modes, the dipolar interaction of the atoms with the incident excitation laser also needs to be taken into account. 

To model the system, we use the following Hamiltonian, that describes the dipole-dipole interactions between the atoms, as well as the interaction of the individual atoms with the incident laser beam:
\begin{eqnarray}
\hat{H}_{total} =  \sum_j \sum_{k}  \left( F_{jk}  \hat{\sigma}_+^j  \hat{\sigma}_-^{k} +F_{kj}  \hat{\sigma}_-^{j}  \hat{\sigma}_+^{k} \right) - \sum_j E_j \hat{\mu}_j \quad . 
\end{eqnarray}

\noindent Here, the first double-summation term describes the dipole-dipole interaction with pairs of atoms, $j$ and $k$, with atomic raising and lowering operators, $\hat{\sigma}_+^j$ and $\hat{\sigma}_-^k$, respectively (the atomic operators are defined as $\hat{\sigma}_+^j  = |1\rangle^j \hspace{0.1cm} {^j}\langle 0|$ and 
$\hat{\sigma}_{-}^j  = |0\rangle^j \hspace{0.1cm} {^j}\langle 1|$). This term in the Hamiltonian is essentially a ``spin" exchange interaction (mediated by vacuum electromagnetic modes) with coupling constants of $F_{jk}$:
\begin{eqnarray}
F_{jk}= F_{kj} &=& -\left(i \frac{\Gamma_a}{2}\right) \left( \frac{3}{8 \pi} \right) \left[ 4 \pi  (1-\cos^2\theta_{jk}) \frac{\sin k_a r_{jk}}{k_a r_{jk}} 
+ 4 \pi (1-3\cos^2\theta_{jk})  \left(\frac{\cos k_a r_{jk}}{(k_a r_{jk})^2} -\frac{\sin k_a r_{jk}}{(k_a r_{jk})^3} \right)   \right] \quad . 
\label{m5}
\end{eqnarray}

\noindent In these coupling constants, the quantity $r_{jk}$ is the distance between the two atoms, and $\theta_{jk}$ is the angle between the atomic dipole moment vector and the separation vector $\vec{r}_{jk}$. The quantity $k_a$ is the wave vector for the photon modes that are energy resonant with the two-level transition: $k_a = \omega_a/c =2 \pi /\lambda_a$ ($\lambda_a$ is the wavelength of the emitted radiation). This dipole-dipole exchange Hamiltonian can be derived by using the interaction of the atomic ensemble with a continuum of radiation modes (i.e., infinite degrees of freedom), and tracing out the radiation coordinates. This derivation is discussed in detail, for example, in Ref.~\cite{ben}. Further details can be found in Appendix~C. 

The second term in Eq.~(7) is the dipolar interaction of the incident laser beam with individual atoms. The quantity $E_j$ is the electric field of the applied laser beam at the position of the $j$th atom and the operators $\hat{\mu}_j$ are the dipole moment operators for the $j$th atom:
\begin{eqnarray}
\hat{\mu}_j \equiv \overbrace{\hat{I} \otimes \hat{I} \otimes  . . . \otimes \hat{I}}^{j-1 \text{ terms}} \otimes \hat{\mu} \otimes \hat{I} \otimes . . . \otimes \hat{I} \quad .  
\end{eqnarray}

\noindent Here, $\hat{I}$ denotes the identity operator for all the other atoms in the ensemble. 

As we mentioned above, in our experiments, the excitation laser is very weak (tens of photons per pulse) to ensure that optical pumping and saturation effects do not come into play. Moreover, the spin exchange term commutes with the collective excitation operator $\hat{J}_{z}=\sum_{j=1}^{N}\hat{\sigma}^{j}_{z}/2$, because
\begin{equation}
    [\hat{J}_{z},\hat{\sigma}^{j}_{+}\hat{\sigma}^{k}_{-}]=0,\forall\; j,k\in\{1,2,\cdots,N\}\quad .
\end{equation}

This allows us to  restrict the problem up to the single excited subspace in our numerical analysis,  with the following basis of states:
\begin{eqnarray}
|000\dots0\rangle, \quad |j \rangle = |100...0 \rangle, \quad  |010...0 \rangle, \quad  |001...0 \rangle,  \quad ... \quad   |000...1 \rangle. 
\end{eqnarray}
The first spin exchange term in $\hat{H}_{total}$ only connects first excited states $|j\rangle$ among themselves whereas the incident laser dipole interaction acts as a perturbation that up to first order only connects the first excited states $|j\rangle$ with the ground state $|000\dots0\rangle$. The perturbative assumption is critical as it reduces the size of the problem from the exponentially large Hilbert space dimension of $2^N$ to $N+1$. With these definitions, we expand the total wavefunction of the atomic ensemble in the above-mentioned basis as:
\begin{eqnarray}
|\psi (t)\rangle = c_0(t) |000...0 \rangle + \sum_{j=1}^{N}c_{j}(t) |j \rangle \quad .
\label{a1} 
\end{eqnarray}

\noindent We proceed perturbatively and take $c_{j}(t)\sim\mathcal{O}(E_{j})$, $\sum_{j=1}^{N} \vert c_{j}(t) \vert^2  << 1$ and $c_0(t) \approx 1$ at all times.  Before the excitation laser beam is applied, the $N$ atom ensemble is in its overall ground state. We, therefore, have the following initial conditions for the  probability amplitudes: $c_0(t=0)=1$ and $c_j(t=0) =0$ for all $j=1...N$. Using the expansion of the wavefunction in Eq.~(12), we next write the Schrodinger's equation using the total Hamiltonian of Eq.~(7):
\begin{eqnarray}
i  \frac{d |\psi(t) \rangle }{dt} & = & \hat{H}_{total} |\psi(t) \rangle \quad  , \nonumber \\
\Rightarrow i \frac{ d c_j(t)}{dt} & = & \sum_k F_{jk} c_k + \Omega_j c_0 \quad .
\label{a4}
\end{eqnarray}

\noindent In above, we have defined the Rabi frequency of the excitation laser at the position of each atom, which is the product of the electric field at that atomic position times the dipole matrix element of the two level transition: $\Omega_j \equiv E_j \mu_{dipole} /\hbar\sim\mathcal{O}(E_{j})$.

We note that Eq.~(13) represents a system of $N$ coupled equations and we numerically solve these as we vary the parameters of the atomic ensemble. Further details can be found in Appendix C. With the probability amplitudes $c_j(t)$ numerically calculated, we then calculate the established dipole moment in the ensemble, which is a measure of the optical power that is absorbed from the laser:
\begin{eqnarray}
P(t) & = &  \langle \psi (t) | \sum_{j=1}^N \hat{\mu}_j  | \psi(t) \rangle \quad . 
\end{eqnarray}

\begin{figure}[h]
\vspace{-1cm}
\begin{center}
%\hspace{3cm}
\includegraphics[width=0.9\textwidth]{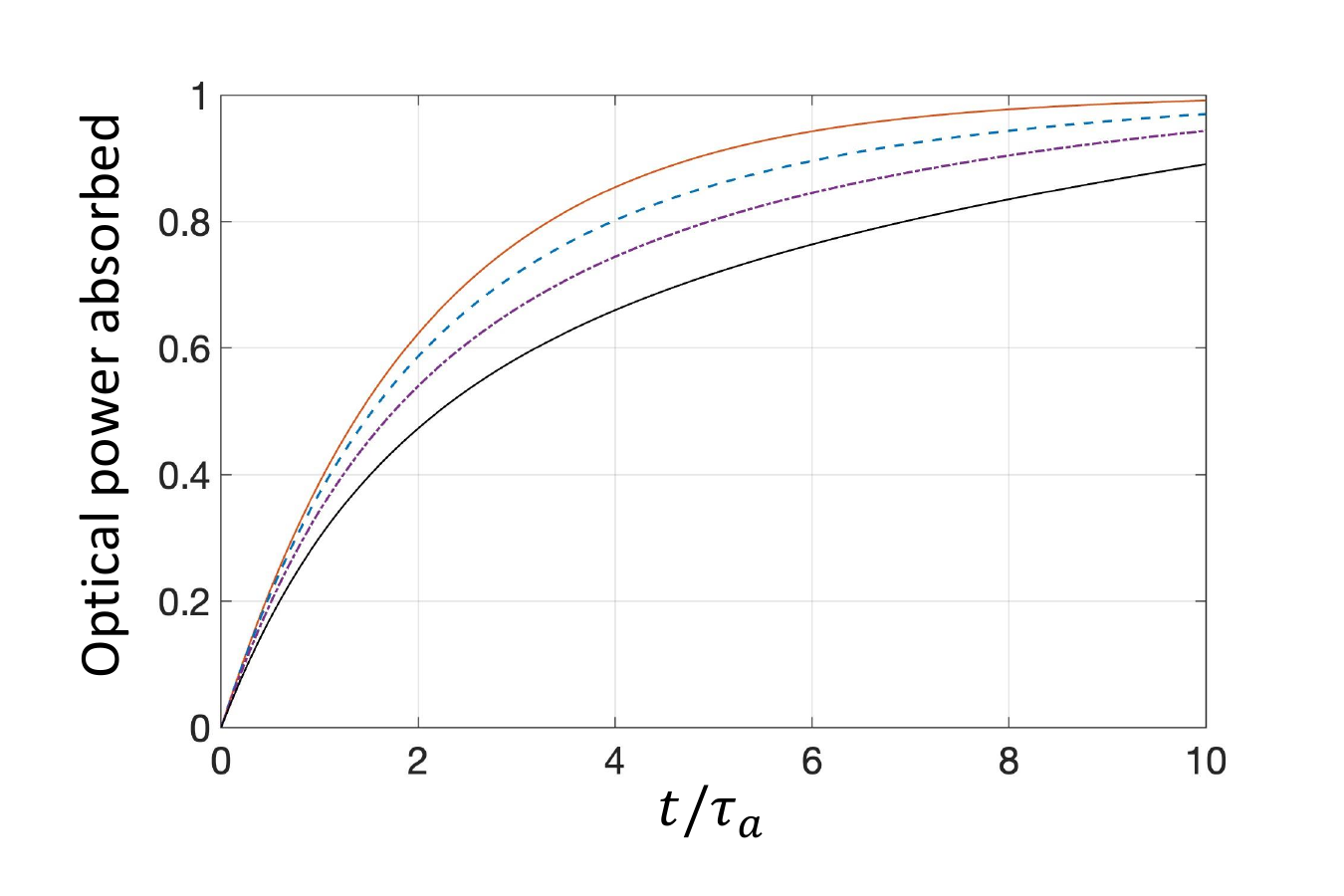}
\vspace{-0.2cm} 
\begin{singlespace}
\caption{\label{scheme} \small Normalized optical power absorbed by the ensemble as a function of time for $N=500$ atoms. Different curves correspond to different volumes of the ensemble. Specifically, the sizes are $50 \lambda_a \times 50 \lambda_a \times 50 \lambda_a$ (solid red), $20 \lambda_a \times 20 \lambda_a \times 20 \lambda_a$ (dashed blue), $15 \lambda_a \times 15 \lambda_a \times 15 \lambda_a$ (dotted-dashed purple), and  $12 \lambda_a \times 12 \lambda_a \times 12 \lambda_a$ (solid black), respectively. When the ensemble is very dilute (solid red curve), the response approaches to what is expected from single-atom physics, with a rise-time of $\tau = 2 \tau_a$. As the volume of the ensemble decreases (and the density increases), the rise-time is considerably increased and subabsorption is observed.   }
\end{singlespace}
\end{center}
\vspace{-0.cm}
\end{figure}

Figure~6 shows sample numerical calculations where we plot the normalized optical power absorbed by the ensemble as a function of the evolution time. Here, to facilitate the simulations, we fix the number of atoms to $N=500$, and change the size of the ensemble so that we obtain a density and optical depth that are comparable to the experimental parameters.  The four curves in Fig.~6 correspond to ensemble sizes of $50 \lambda_a \times 50 \lambda_a \times 50 \lambda_a$ (solid red), $20 \lambda_a \times 20 \lambda_a \times 20 \lambda_a$ (dashed blue), $15 \lambda_a \times 15 \lambda_a \times 15 \lambda_a$ (dotted-dashed purple), and  $12 \lambda_a \times 12 \lambda_a \times 12 \lambda_a$ (solid black), respectively. When the ensemble is very dilute (solid red curve), the response approaches to what is expected from single-atom response, with a rise-time of $\tau = 2 \tau_a$. As the volume of the ensemble decreases (and the density and the optical depth increases), the absorption rise-time is considerably increased and subabsorption is observed. We find that the overall behavior in these evolution curves is similar to collective decay in dilute samples at a low optical depth \cite{davidexp,yavuz_coherence,rudhy}. In collective spontaneous emission in this regime, there may be slight early-time superradiance, but the collective physics is largely dominated by subradiant decay observed at later times. Similarly, as shown in Fig.~6, while there may be slight early-time superabsorption, the main physical effect is subabsorption which dominates the dynamics. 

Using this model, the amount of subabsorption that we predict is far larger than what we observe experimentally. We believe that this is due to incoherent processes, such as radiation trapping and van der Waals dephasing \cite{haroche}, reducing the established dipole-dipole correlations across the ensemble as the density and the optical depth is increased. This assumption is consistent with our recent experiments, where we observed that the subradiance in dilute clouds increased up to a certain optical depth and then displayed a clear reduction \cite{davidexp}. Since then, we have obtained more data on subradiance in dilute clouds for different parameters, which also displayed a similar behavior \cite{david_inprep}. 

We believe it is difficult to accurately model such incoherent processes in our system since the physics of collective coupling in dilute clouds rely on many-atom correlations. We leave a detailed study of a quantitative model of such incoherent processes in dilute clouds for future work. Here, we instead use the following heuristic approach. We assume that there is a density dependent dephasing with a rate $\gamma_{DD} = \beta n$, that reduces the established dipole-dipole correlations, by reducing the coupling constants of the exchange Hamiltonian in a manner similar to a Lorentzian lineshape:
\begin{eqnarray}
    F_{jk} \rightarrow \frac{1}{1+\left( \gamma_{DD}/\Gamma_a \right)^2} F_{jk} \quad . 
\end{eqnarray}

\noindent We then adjust the density dependent broadening coefficient $\beta$ so that the subabsorption peak that is observed in numerical simulations matches the experimental data of Fig.~4. Figure~7 shows the numerically calculated absorption risetime $\tau$ (again normalized to $2 \tau_a$) as a function of the optical depth of the ensemble, as the broadening coefficient is changed from $\beta/2 \pi =0 $ to $\beta/2 \pi  = 9 \times 10^{-5}$ Hz~cm$^3$. As the dipole-dipole broadening coefficient is increased, the observed subabsorption decreases and peaks at lower values of the optical depth. The best agreement with the experimental results is obtained for a broadening coefficient of $\beta/2 \pi  = 4.9 \times 10^{-5}$ Hz~cm$^3$. The numerical calculation for this specific value of the dephasing coefficient is plotted in Fig.~4(b) and is in reasonable agreement with the experimental data. 

We note that using this single fitting parameter, the experimental data shows a sharper feature for the subabsorption signal as a function of the optical depth of the ensemble compared to the numerical calculations. We speculate that this may be due to (1) an additional dephasing mechanism in the experiment that becomes dominant at low densities and decoheres dipole-dipole correlations, or (2) the multi-level hyperfine structure of $^{87}$Rb atoms, which is ignored in the numerical calculations.

\begin{figure}[h]
\vspace{0cm}
\begin{center}
%\hspace{3cm}
\includegraphics[width=1.1\textwidth]{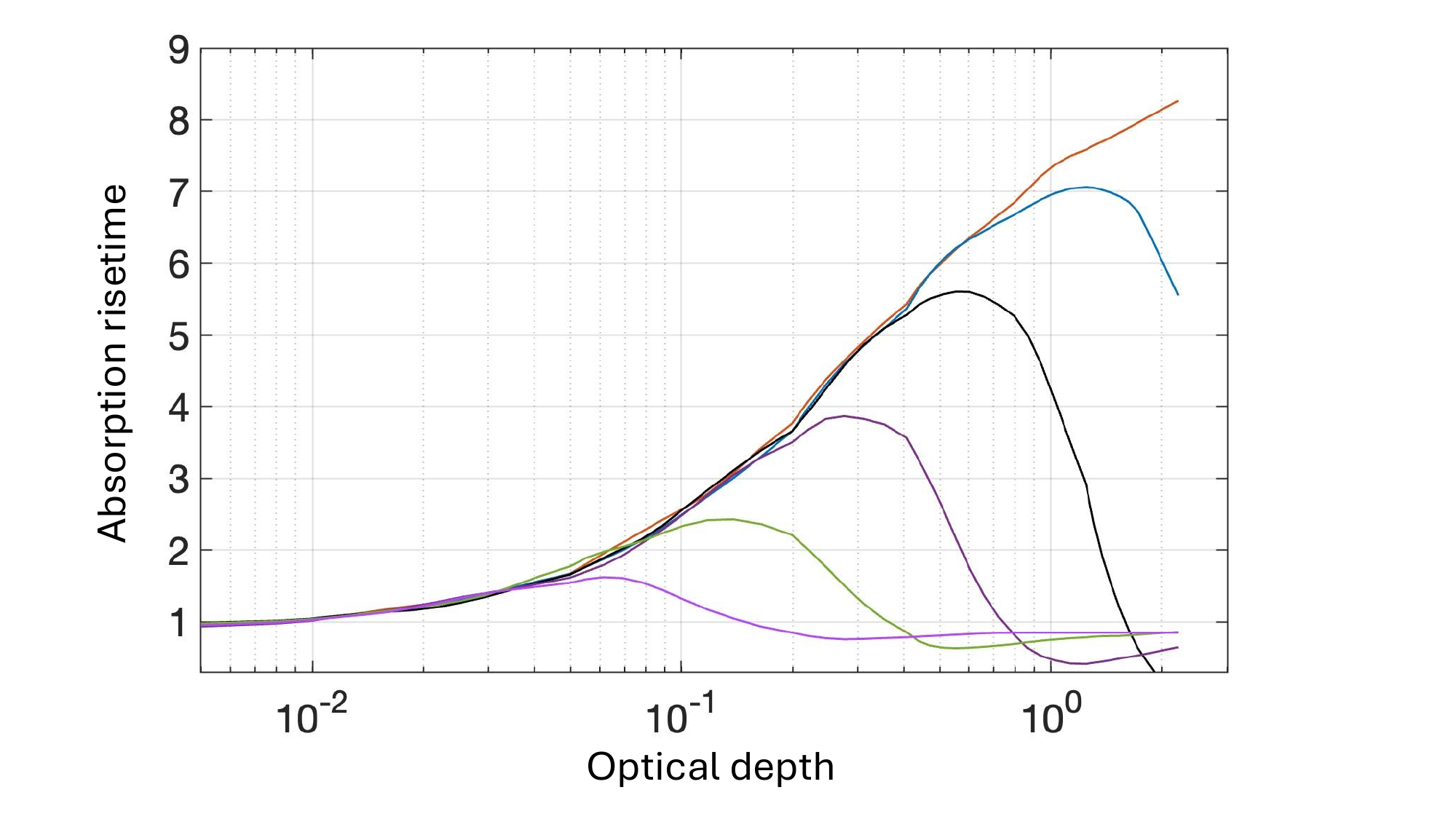}
\vspace{-0.7cm} 
\begin{singlespace}
\caption{\label{scheme} \small Numerically calculated absorption rise-time $\tau$ (normalized to $2 \tau_a$) as a function of the optical depth of the ensemble, for $\beta/2 \pi =0 $ (solid orange curve), $\beta/2 \pi  = 9 \times 10^{-7}$ Hz~cm$^3$ (solid blue), $\beta/2 \pi  = 2.8 \times 10^{-6}$ Hz~cm$^3$ (solid black), $\beta/2 \pi  = 9 \times 10^{-6}$ Hz~cm$^3$ (solid purple), $\beta/2 \pi  = 2.8 \times 10^{-5}$ Hz~cm$^3$ (solid green), and $\beta/2 \pi  = 9 \times 10^{-5}$ Hz~cm$^3$ (solid teal), respectively. As the density-dependent dephasing coefficient is increased, the observed subradiance decreases and peaks at lower values of the optical depth. The best agreement with the experimental data using this single adjustable parameter is plotted in Fig.~4(b).}
\end{singlespace}
\end{center}
\vspace{-0.cm}
\end{figure}

We also note that the inferred density-dependent dephasing coefficient that provides the best agreement with the experiment, $\beta/2 \pi  = 4.9 \times 10^{-5}$ Hz~cm$^3$, is more than two orders of magnitude larger than the well-known dipole-dipole line broadening coefficient in $^{87}$Rb \cite{adams_dipole}. We believe this discrepancy is not unexpected. Multi-atom correlations are expected to dephase more quickly compared to the straightforward broadening of the single-atom lineshape. To first order, we would expect the multi-atom dephasing rate to scale linearly by the average number of correlated atoms. This suggests that the average number of correlated atoms is high, which is consistent with the sensitive nature of the subabsorption signal to the motional dephasing,  and therefore to the atomic temperature. This is also consistent with the spatial coherence of light in collective decay that we observed in our recent experiment \cite{davidexp}, which suggested that states correlated across the whole ensemble significantly contributed to the collective coupling in dilute samples. 

\section{Conclusions}

In conclusion, we have predicted and experimentally demonstrated a new type of collective coupling effect where a disordered atomic ensemble absorbs light with a longer rise-time (i.e., at a slower rate) than what is dictated by the excited state lifetime. This effect, which we name subabsorption, can be viewed as the absorptive analog of subradiance. We performed our experiment using a dilute ensemble of laser-cooled $^{87}$Rb atoms with a low optical depth, and time-resolving the absorption of a weak (tens of photons per pulse) resonant laser beam. We have investigated subabsorption as a function of the optical depth of the ensemble, and demonstrated that this effect is observed in dilute clouds with a low optical depth. 

The physics of subabsorption is  similar to recently observed subradiance and spatial coherence of light in collective spontaneous emission \cite{davidexp,yavuz_coherence} in dilute atomic ensembles. The collective interaction relies on establishing dipole-dipole correlations over many atoms; i.e., the interactions are not dominated by the nearest neighbors. As a result,  subabsorption is highly susceptible to motional dephasing: even a temperature increase of 60 $\mu$K (i.e., motional dephasing of $\sim \lambda_a/100$ in an atomic lifetime) is enough to completely extinguish the subabsorption signal. We have also presented a theoretical model whose results are in reasonable agreement with the experimental observations. The model uses a single density-dependent dephasing rate of the long-range dipole-dipole correlations and experiment-theory comparison indicates a dephasing coefficient of $\beta/2 \pi  = 4.9 \times 10^{-5}$ Hz~cm$^3$.    

A quantitative model for the observed dephasing as well as further investigations of the experiment-theory comparison is one clear future direction. In this work we have modeled the dephasing heuristically by assuming a single density dependent rate, $\gamma_{DD} = \beta n$, and suppressing the dipole-dipole interaction coefficients in the manner described in Eq.~(15). Furthermore, our numerical model made a number of simplifying assumptions. For example, we ignored the multi-level hyperfine structure of $^{87}$Rb atoms and assumed an ensemble of two level atoms. Our numerical model also ignored various imperfections in the experiment such as the Doppler broadening of the resonance, frequency fluctuations of the lasers, and the transverse spatial profile of the atomic cloud and the excitation laser pulse.  In the near future, moving beyond these approximations may provide better insight into the experimental results. 

Another future direction is to simultaneously study the radiated light from the atomic ensemble in a direction orthogonal to the laser propagation direction, while also measuring subabsorption of the incident laser beam. The same multi-atom correlations that produce the subabsorption signal will also have an effect on the spatial structure and coherence properties of the emitted radiation. Studying quantum statistics of the emitted radiation, as well as the photon statistics of the excitation laser after it experiences suababsorption through the cloud would also be very interesting. Because multi-atom correlations play a role, deviation from the Poissonian statistics of the coherent state is expected \cite{mandel_wolf}. Using homodyne detection with a strong local oscillator, the two quadratures of the excitation laser can be measured and deviations from the coherent state can be quantified. 

Much recent work in collective decay has focused on ordered atomic arrays \cite{yelin,bloch,kimble}. It is known that collective (superradiance and subradiance) decay rates from such arrays depend sensitively on the array parameters \cite{kimble,rudhy}. Similarly, subabsorption can be studied both experimentally and theoretically using atoms trapped in one, two, or three dimensional arrays. Recent advances in neutral atom quantum computing have achieved atomic arrays that can trap 1000 atoms or more \cite{mark_review,browaeys_large,lukin_logical}. Similar to the experiments that we described above, absorption rise-time for a weak resonant laser beam propagating through such arrays can be experimentally investigated. For example, absorption rise-times will depend sensitively on the array parameters, such as the number of atoms in the array and the spacing between the atoms. 

We also note that interaction of light with an ensemble of radiators and collective coupling is important in many physical processes that we observe in nature. For example, superradiance is now believed to be relevant even in many astrophysical phenomenon, including black-hole formation \cite{rosa,witek,sarmah,teo,zhou}. We believe it is plausible that absorptive counterparts of collective effects, superabsorption and  subabsorption, may also be significant in a wide range of physical systems. An exciting future direction is to explore universal parameter regimes where subabsorption comes into play, and investigate its relevance in various length scales, including in astrophysical objects.

\section{Acknowledgments}

We thank Akbar Safari for helpful discussions. This work was supported by the National Science Foundation (NSF) Grant No. 2016136 for the QLCI center Hybrid Quantum Architectures and Networks (HQAN), NSF Grant No. 2308818 from the AMO-Experiment program, and also by the University of Wisconsin-Madison, through the Vilas Associates award.

\section{Appendix A: experimental details and the data fitting procedure}

The experiment is performed inside an ultrahigh vacuum chamber which is kept at a base pressure of $\sim 10^{-9}$~torr. To form the $^{87}$Rb MOT, we use three counter-propagating beam pairs that are locked to the $F=2 \rightarrow F'=3$ cycling transition in the D2 line (with a transition wavelength of 780 nm). Two beam-pairs each have an optical power of about 40 mW and a beam radius of 3 cm. Due to space constraints, the third beam pair is not orthogonal to the other two, and has a smaller size, with a beam radius of 5 mm, and an optical power of 5 mW. The MOT lasers are produced by a custom-built external-cavity diode laser (ECDL) whose output is amplified by a semiconductor tapered amplifier, before being split into counter-propagating beam pairs. Further details regarding our laser system can be found in our prior publications \cite{dipto,davidexp}. The MOT lasers are overlapped with a hyperfine repumping beam, which is generated by a separate ECDL locked to the $F=2 \rightarrow F'=2$ transition with an optical power of about 1 mW.  

We load the atoms to the MOT from the background vapor in the chamber for about 100 ms. For the last 20 ms of loading, we detune the MOT lasers by about $8 \Gamma_a$ from the cycling transition and reduce their intensity by about an order of magnitude to achieve efficient sub-Doppler cooling. At the end of the MOT loading cycle, we typically trap $10^5$ atoms, within a radius of 0.26 mm. The atomic temperature is about $40$~$\mu$K, which is measured by monitoring the free expansion of the cloud using an electron-multiplying CCD camera. During the final few ms of the MOT loading cycle, we turn-off the hyperfine repumper beam. As a result, the atoms are optically pumped into the $F=2$ ground level at the end of the MOT loading cycle. 

The excitation laser beam is obtained by picking off a small portion of the MOT laser and tuning it to the cycling resonance with the use of a fast acousto-optic modulator (AOM). The excitation beam is loosely focused on the atomic cloud with a long focal length lens, and has a spatial size that is comparable to the size of the ensemble.  After passing through the atomic ensemble, the excitation laser pulse is detected using a single-photon counting module (SPCM).  The output of the SPCM is measured on an oscilloscope, which gives approximately 4 ns time-resolution for the intensity of the pulse.  Due to the low optical depth of the cloud and the long data acquisition times for the experiment, it is essential to account for laser intensity and alignment drift.  To this end, rather than taking two separate series of measurements with and without the atomic cloud, we toggle the cloud on and off between each experimental cycle ($\sim 100$ ms), which we perform by disabling the magnetic quadrupole field for the MOT.  We keep all other parameters the same, including laser pulse sequences, which allows us to make continuous measurements of both $I_{input}$ and $I_{output}$ while minimizing the effects of long-term drift.  

To ensure that saturation and optical pumping effects do not play a role in our experiment, we adjust the excitation laser light intensity such that the number of photons in the excitation laser pulse is at the tens of photons level over the $8\tau_{a}$ data taking and curve fitting duration, which is much smaller than the number of atoms in the ensemble. As a result, the experiment is deeply in the weak excitation regime and the Rabi frequency of the excitation laser pulse is much smaller than the excited state decay rate, $\Omega << \Gamma_a$. We have also experimentally verified that the experiment is in the weak-excitation regime; when the intensity of the incident laser pulse is increased by an order of magnitude, the experimental results are qualitatively unchanged. 

We compute the absorption rise-time, $\tau$, and its uncertainty as follows:  We measure and average $I_{input}$ and $I_{output}$ over $\sim 10^5$ experimental cycles.  We assume Poissonian photon statistics and calculate $\sigma (t)$ at each time step and the corresponding uncertainty. The absorption rise-time $\tau$ is then calculated from the best fit curve of the form exponential rise between $\tau_a < t < 8 \tau_a$. We note that there is a slight experimental uncertainty (at the level of at most 5\%) in the initial [$\sigma (t=t_a)$] and the steady-state ($\sigma_{ss}$) values of the optical depth. As a result, while finding the best exponential fit, we allow these values to vary slightly, which we find gives a better overall fit.

To calculate the uncertainty in the rise-time $\tau$, we use a Monte-Carlo method and for each data point in  $\sigma (t)$ add a randomly sampled value from its uncertainty distribution.  We then calculate a new best fit curve and a new value of the rise-time $\tau$.  We repeat this process $\sim 10^4$ times until we have a large sample of best fit rise times and then calculate its standard deviation, which we use as the uncertainty in $\tau$.  We note that this fitting method gives a reduced chi-squared value of close to 1.

\section{Appendix B: details on Maxwell-Bloch simulations}

To calculate the non-interacting gas response, we solve the partial differential equations for $\rho_{00} (z,t')$, $\rho_{11} (z,t')$, $\rho_{01} (z,t')$, and $\Omega (z,t')$ [Eqs.~(3) and (4)] numerically on a space-time grid using the method of lines. Starting at $z=0$, we first solve the time-domain density matrix equations for the atomic ensemble, Eq.~(3), numerically using fourth-order Runge-Kutta with the initial condition that all the atoms start in the ground state. With the density matrix elements numerically calculated, we then use the atomic coherence $\rho_{01}$, to propagate the laser pulse to the next spatial point, $z \rightarrow z + \Delta z$ using Eq.~(4). This procedure is then repeated until the excitation laser reaches to the end of the atomic ensemble. 

A sample numerical calculation for an atomic ensemble with a steady-state optical depth of $\sigma_{ss}=0.5$ is shown in Fig.~8. Here, we plot the normalized input laser pulse to the ensemble, $I_{input}$, the output pulse at the end of the cloud, $I_{output}$, as well as the established coherence at the beginning of the atomic cloud, $\vert \rho_{01} (z=0, t) \vert$, and the optical depth as a function of time, $\sigma (t)$. The plotted input and output pulses have the same normalization. The optical depth as a function of time is estimated by using the numerically-calculated output pulse and by relating the input and output intensity as described in Eq.~(1).

\begin{figure}[p]
\vspace{0cm}
\begin{center}
%\hspace{3cm}
\includegraphics[width=0.6\textwidth]{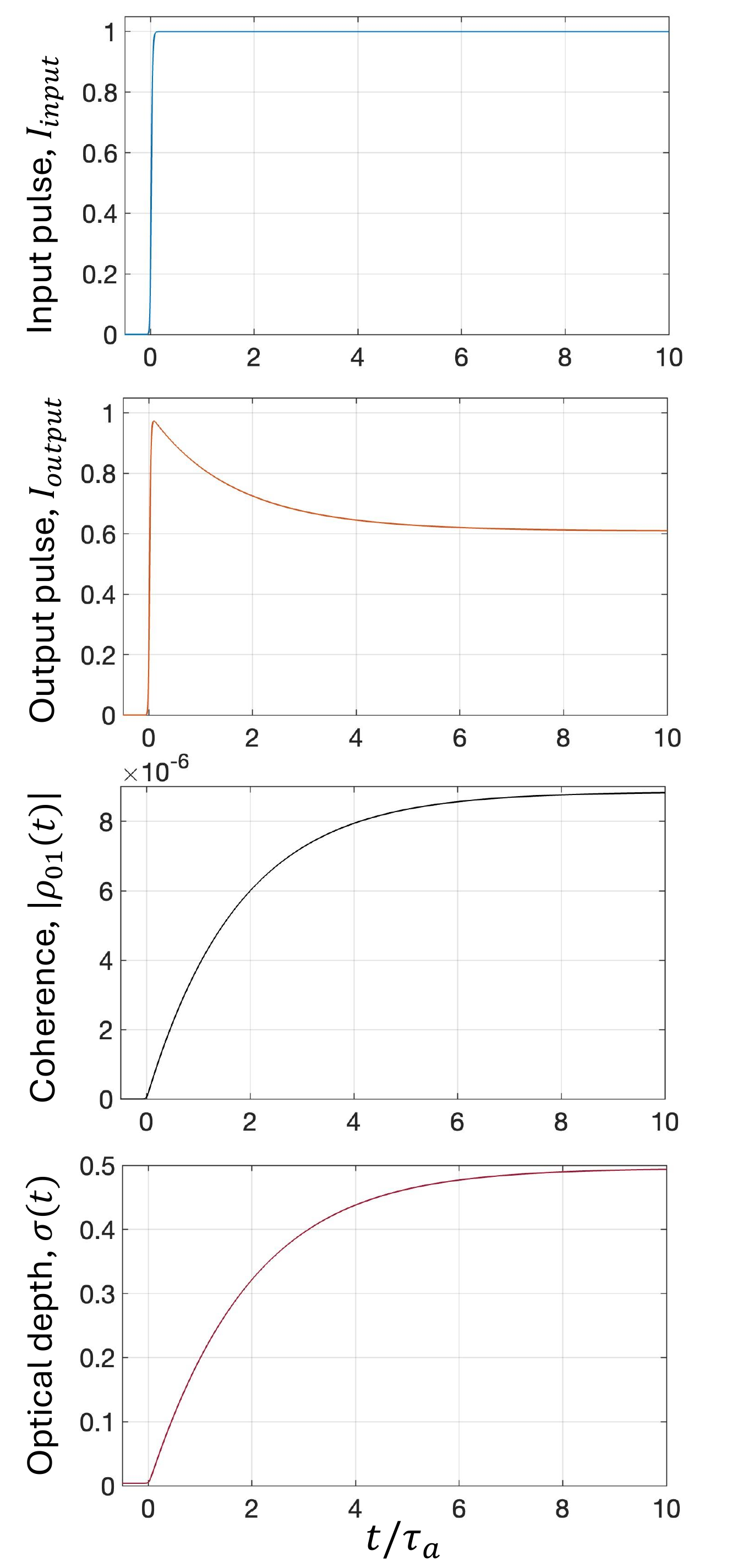}
\vspace{-0.2cm} 
\begin{singlespace}
\caption{\label{scheme} \small The normalized input laser pulse to the ensemble, $I_{input}$, the output pulse, $I_{output}$, the established coherence at the beginning of the atomic cloud, $\vert \rho_{01} (z=0, t) \vert$, and the optical depth as a function of time $\sigma (t)$, for a sample Maxwell-Bloch numerical simulation. This simulation is performed for an atomic cloud with a steady-state optical depth of $\sigma_{ss}=0.5$.}
\end{singlespace}
\end{center}
\vspace{-0.cm}
\end{figure}

The solid line in Fig.~4(a) is calculated by performing simulations that are similar to those shown in Fig.~8, but with varying optical depth. For each optical depth, the absorption rise-time is then estimated using an exponential fit to the optical depth as a function of time, $\sigma(t)$. The quality of the exponential fits to the numerical results is very good; as a result, the error bars on the calculated rise-time are negligible. Therefore, these error bars are not visible in the plot of Fig.~4(a) (the solid red curve).

\section{Appendix C: details on the collective simulations}

 The atomic-only exchange Hamiltonian which is given in the first term on the right hand side of Eq.~(7) can be derived from the full Hamiltonian that describes the interaction of the atomic ensemble with a continuum of radiation modes: 
	\begin{eqnarray}
	    \hat{H}=\sum_{j=1}^{N}\frac{\omega_{a}\hat{\sigma}^{j}_{z}}{2}+\sum_{\vec{k},\vec{\epsilon}}\omega_{\vec{k},
	    \vec{\epsilon}}\left(\hat{a}^{\dagger}_{\vec{k},\vec{\epsilon}}\hat{a}_{\vec{k},\vec{\epsilon}}+\frac{1}{2}\right)
	    -\sum_{j=1}^{N}\sum_{\vec{k},\vec{\epsilon}}(g^{*}_{\vec{k},\vec{\epsilon}}\,e^{i \vec{k}\cdot\vec{r}_{j}}\hat{\sigma}^{j}_{+}\hat{a}_{\vec{k},\vec{\epsilon}}+g_{\vec{k},\vec{\epsilon}}\,e^{-i \vec{k}\cdot\vec{r}_{j}}\hat{\sigma}^{j}_{-}\hat{a}^{\dagger}_{\vec{k},\vec{\epsilon}}) \quad , 
	\end{eqnarray}
	where 
\begin{eqnarray}
\hat{\sigma}_z^j & =& |1\rangle^j \hspace{0.1cm} {^j}\langle 1|-|0\rangle^j \hspace{0.1cm} {^j}\langle 0| \quad , \nonumber \\
\hat{\sigma}_+^j & =& |1\rangle^j \hspace{0.1cm} {^j}\langle 0| \quad , \nonumber \\
\hat{\sigma}_{-}^j & =& |0\rangle^j \hspace{0.1cm} {^j}\langle 1| \quad ,
\label{m2}
\end{eqnarray}

\noindent are the atomic spin operators for the $j$th atom with energy eigenstates $|0 \rangle^j$ and $|1 \rangle^j$, respectively.  The operators $\hat{a}_{\vec{k},\vec{\epsilon}}$ and $\hat{a}^{\dagger}_{\vec{k},\vec{\epsilon}}$, are the photon annihilation and creation operators for a radiation mode with wave-vector $\vec{k}$ and polarization $\vec{\epsilon}$. The well-known Dicke limit can be obtained from the above Hamiltonian when the size of the sample is small compared to the radiation wavelength set by the relevant $k$, i.e., $\vec{k}\cdot\vec{r}_{j}\rightarrow 0,\forall\,\,0\le j\le N$ ($\vec{r}_{j}$ is the position of the $j$th atom). Using the Born-Markov approximation and tracing out the radiation modes, we can reduce the above full-Hamiltonian to the atomic-only exchange Hamiltonian of Eq.~(7). This derivation is discussed in detail, for example, in Ref.~\cite{ben}.

We note that Eq.~(13) represents a system of $N$ coupled equations where $N$ is the number of atoms in the ensemble. To solve these equations numerically, it is useful to define the following vectors:
\begin{eqnarray}
\vec{c}(t) \equiv  \left( \begin{array}{c} 
c_1(t) \\
c_2(t) \\
. \\
. \\
. \\
c_N (t) \end{array} \right)   \quad , \quad  
\vec{\Omega} \equiv -i  \left( \begin{array}{c} 
\Omega_1 \\
\Omega_2 \\
. \\
. \\
. \\
\Omega_N \end{array} \right)
\quad .
\end{eqnarray}

\noindent With these definitions, the coupled equations of Eq.~(13) can be written in the following vector-matrix notation in the perturbative limit, $c_0(t) \approx 1$:
\begin{eqnarray}
\frac{d \vec{c}}{dt} = - \bar{\bar{H}} \cdot \vec{c} + \vec{\Omega} \quad .  
\end{eqnarray} 

\noindent Here, $\bar{\bar{H}}$ is an $N \times N$ matrix, whose entries are the dipole-dipole coupling constants (up to a phase factor) that are given in Eq.~(8), i.e., $\bar{\bar{H}}_{jk} = i F_{jk}$. We now need to solve Eq.~(19) with the initial condition that the system starts in its overall ground state and $\vec{c}(t=0) = \vec{0}$. When the inhomogeneous driving term is time-invariant, $\vec{\Omega}(t)= \vec{\Omega}(t=0)$, this solution can be written analytically and is:
\begin{eqnarray}
\vec{c}(t) = \exp{\left( - \bar{\bar{H}} t \right) } \cdot \bar{\bar{H}}^{-1} \cdot \exp{\left(  \bar{\bar{H}} t \right) } \cdot \vec{\Omega} - \exp{\left( - \bar{\bar{H}} t \right) } \cdot \bar{\bar{H}}^{-1} \cdot \vec{\Omega} \quad . 
\end{eqnarray}

\noindent Here, $\exp{( -\bar{\bar{H}} t )}$ refers to matrix exponentiation of the $N \times N$ matrix. In the numerical results that we presented above, we calculate the state evolution numerically using Eq.~(20) as we vary the parameters of the ensemble. We take the excitation laser to be propagating along $z$ and polarized orthogonal to the propagation direction. We, therefore, use the electric-field for the excitation laser beam of the form $E_j = \hat{x} E_0 \exp{(i k_a z_j)}$ (the quantity $z_j$ is the $z$ coordinate of the $j$th atom), calculate the state evolution, and find the established dipole moment using the probability amplitudes by evaluating Eq.~(14). With the dipole moment $P(t)$ calculated, we then find the absorption rise-time by finding the best exponential fit to the established dipole moment as a function of time. These numerical results are presented in Figs.~6 and 7 of above, as well as in Fig.~4(b) which is the calculation that gives best agreement with the experimental results. 

\begin{figure}[h]
\vspace{0cm}
\begin{center}
%\hspace{3cm}
\includegraphics[width=1.1\textwidth]{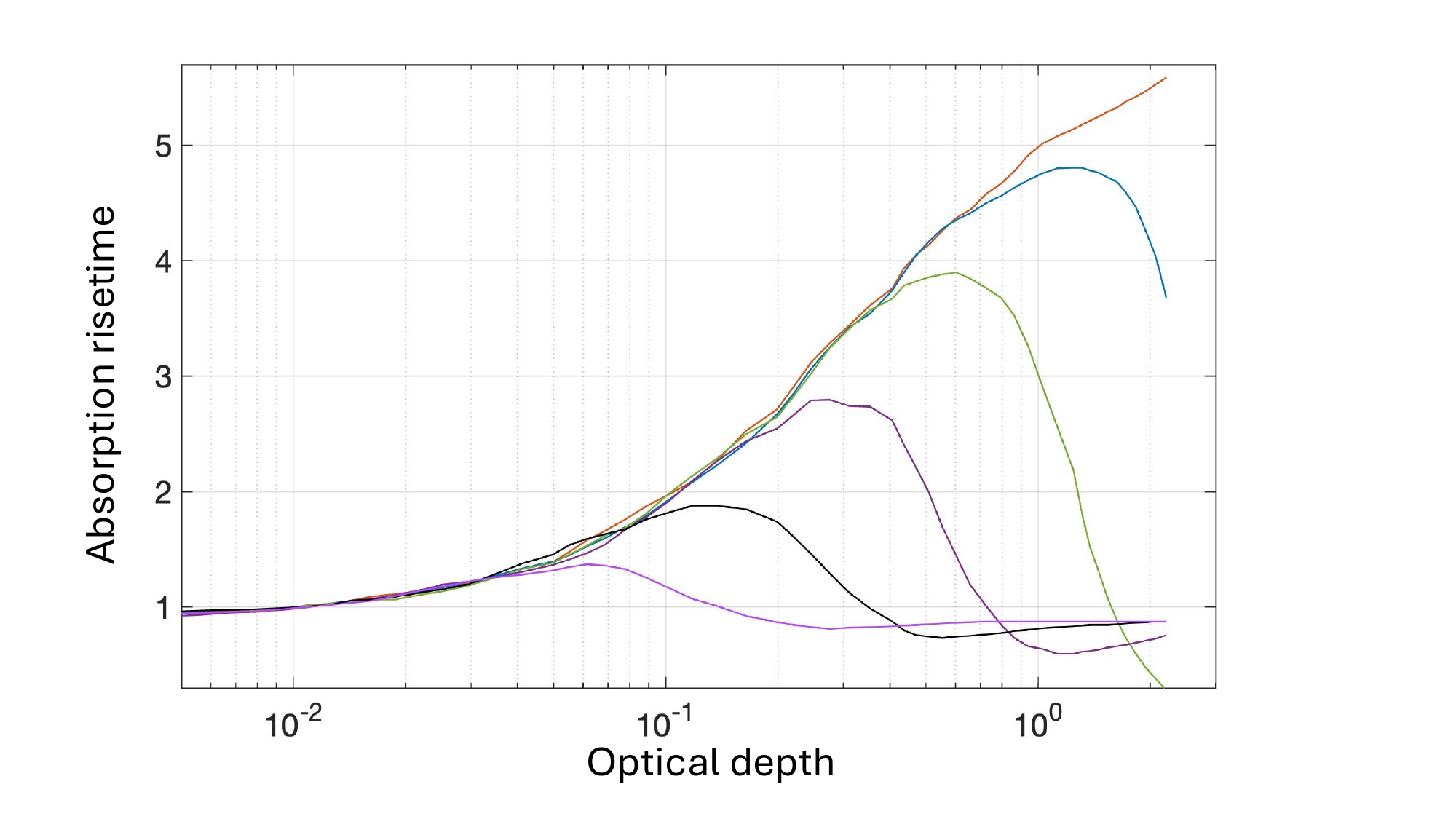}
\vspace{-0.7cm} 
\begin{singlespace}
\caption{\label{scheme} \small Numerically calculated absorption risetime $\tau$ (normalized to $2 \tau_a$) as a function of the optical depth of the ensemble, for $\beta/2 \pi =0 $ (solid orange curve), $\beta/2 \pi  = 9 \times 10^{-7}$ Hz~cm$^3$ (solid blue), $\beta/2 \pi  = 2.8 \times 10^{-6}$ Hz~cm$^3$ (solid black), $\beta/2 \pi  = 9 \times 10^{-6}$ Hz~cm$^3$ (solid purple), $\beta/2 \pi  = 2.8 \times 10^{-5}$ Hz~cm$^3$ (solid green), and $\beta/2 \pi  = 9 \times 10^{-5}$ Hz~cm$^3$ (solid teal), respectively. This is exactly the same calculation as in Fig.~7, while keeping the angle between the dipole moment vector and position vector between two atoms, $\theta_{jk}$, versus setting this angle to $\theta_{jk}=0$ (i.e., roughly contrasting scalar versus vectorial model). While there are quantitative differences, qualitatively the results of vectorial and scalar simulations are similar.}
\end{singlespace}
\end{center}
\vspace{-0.5cm}
\end{figure}

Using our numerical model, we have investigated if the vectorial nature of the problem changes the results significantly. For example, in Fig.~9, we do exactly the same calculation as in Fig.~7, while keeping the angle between the dipole moment vector and position vector between two atoms, $\theta_{jk}$, versus setting this angle to $\theta_{jk}=0$ (i.e., roughly contrasting scalar versus vectorial model). Comparing Figs.~7 and 9, we find that while there are quantitative differences, qualitatively the results are similar.

\begin{figure}[h]
\vspace{0cm}
\begin{center}
%\hspace{3cm}
\includegraphics[width=0.9\textwidth]{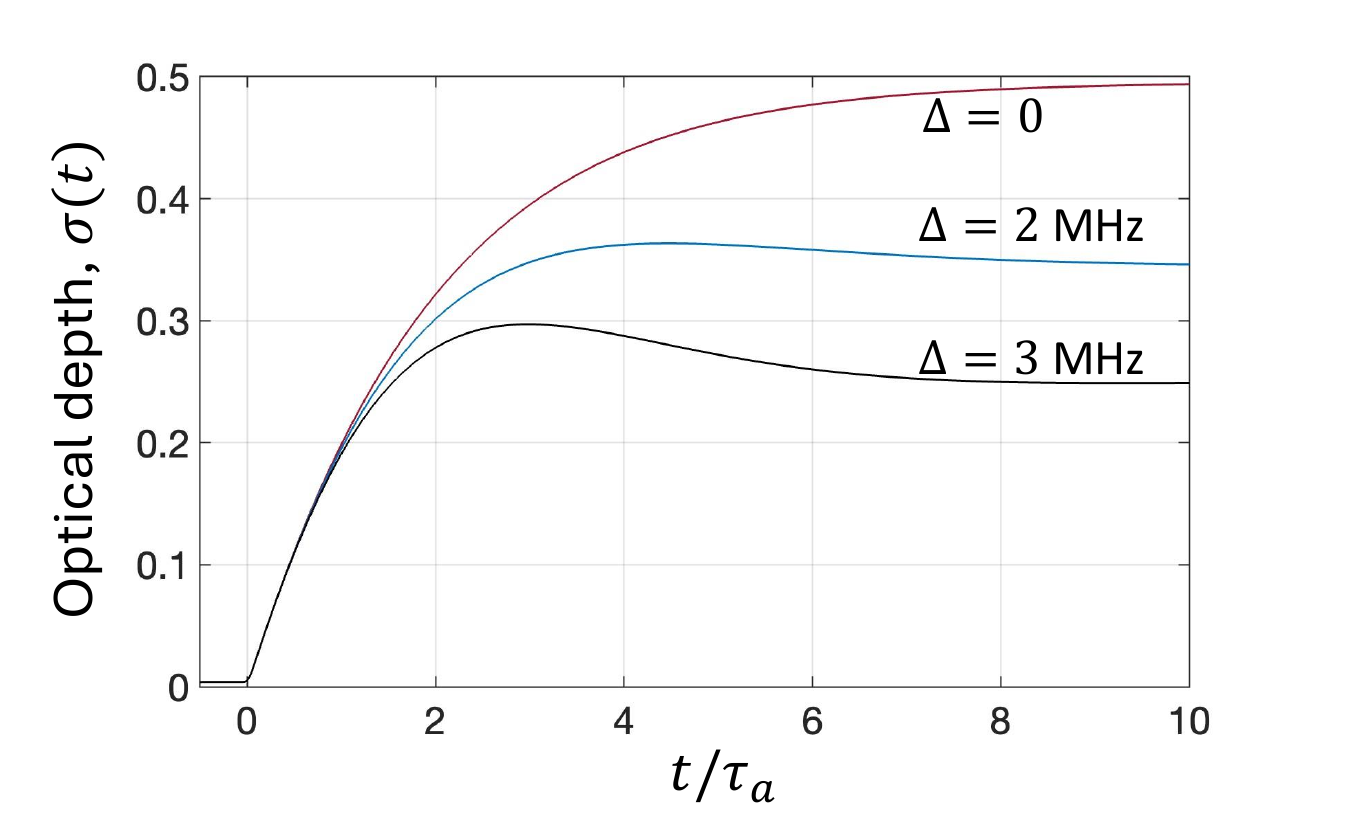}
\vspace{-0.2cm} 
\begin{singlespace}
\caption{\label{scheme} \small Optical depth as a function of time, $\sigma (t)$, for three values of the frequency detuning, $\Delta = 0 $, $\Delta = 2 $~MHz (i.e., detuned by $\Gamma_a/3$), and $\Delta = 3 $~MHz (detuned by $\Gamma_a/2$), respectively. These results are obtained using the non-interacting gas Maxwell-Bloch propagation code. Here, we use an atomic ensemble with a steady-state optical depth of $\sigma_{ss}=0.5$ for on-resonance light (i.e., exactly the same optical depth as the simulations that are displayed in Fig.~8). The shape of the curves, and therefore, the inferred absorption rise-times change significantly, even when the frequency detuning is significantly smaller than the linewidth of the transition. }
\end{singlespace}
\end{center}
\vspace{-0.cm}
\end{figure}

\section{Appendix D: dependence of the absorption rise-time on laser frequency}

A first iteration of the experiment that we reported here failed, which we later found out was primarily due to the long-time-scale frequency drifts of the excitation laser. This is because of the sensitive nature of the optical depth curves, $\sigma(t)$, to the frequency difference (detuning) between the excitation laser and the atomic transition, $\Delta = \omega_a -\omega_{laser}$. Figure~10 shows the numerical Maxwell-Bloch simulations that show this effect for a representative set of parameters. Here, we use an atomic ensemble with a steady-state optical depth of $\sigma_{ss}=0.5$ for on-resonance light (i.e., exactly the same optical depth as the simulations that are displayed in Fig.~8). In Fig.~10, we plot $\sigma (t)$ for three values of the detuning, $\Delta = 0 $, $\Delta = 2 $~MHz (i.e., detuned by $\Gamma_a/3$), and $\Delta = 3 $~MHz (detuned by $\Gamma_a/2$), respectively. The shapes of the curves, and therefore, the inferred absorption rise-times change significantly, even when the frequency detuning is significantly smaller than the linewidth of the transition. 

\begin{figure}[h]
\vspace{0cm}
\begin{center}
%\hspace{3cm}
\includegraphics[width=0.85\textwidth]{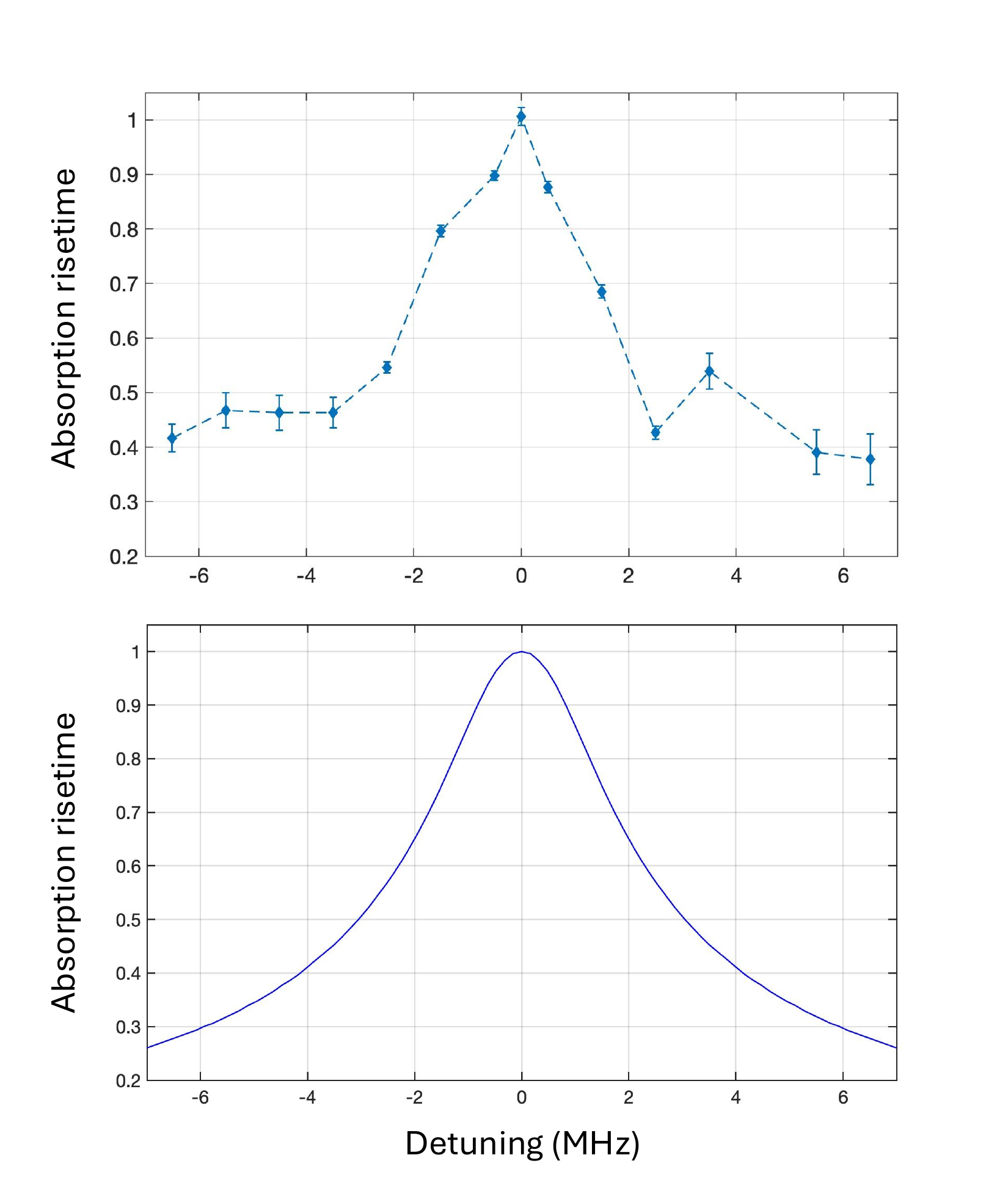}
\vspace{-0.2cm} 
\begin{singlespace}
\caption{\label{scheme} \small The top plot is the experimentally-measured absorption rise-time, $\tau$ as the frequency detuning of the excitation laser, $\Delta$, is varied. For this experiment, we use an atomic ensemble with a steady-state optical depth of $\sigma_{ss} \approx 1$. For $\Delta \approx \Gamma_a$, the rise-time gets shorter by as much as 60~\%. The bottom plot is the calculated rise-time as a function of frequency detuning using the Maxwell-Bloch propagation code. There is reasonable agreement between the experimental results and the calculation. }
\end{singlespace}
\end{center}
\vspace{-0.cm}
\end{figure}

The main issue with our earlier iteration of the experiment was the absolute frequency drifts of the excitation laser. These frequency drifts were of order of 1~MHz, over the long time scales ($\sim$1 day) of data taking ($>10^5$ experimental cycles). As we mentioned above, a large number of experimental cycles is required to obtain good statistics, especially when the optical depth of the ensemble is low. We resolved the frequency drift issue by improving the linewidth of the excitation laser and also the long-term frequency stability of the saturated-absorption lock.

Once we fixed the long-term frequency drift of the excitation laser, we  experimentally investigated the effect of the detuning to the absorption rise-time. In Fig.~11, we experimentally measure the absorption rise-time as the detuning of the excitation laser, $\Delta = \omega_a - \omega_{laser}$, is varied. For this data set, we use an atomic ensemble with a steady-state optical depth of $\sigma_{ss} \approx 1$. For $\Delta \approx \Gamma_a$, the rise-time gets shorter by as much as 60~\%. The bottom plot in Fig.~11 is the calculated rise-time as a function of frequency detuning using the Maxwell-Bloch propagation code (i.e., taking the ensemble as a non-interacting gas). There is reasonable agreement between the experimental results and the calculation. 

The subabsorption results that we presented above in Figs.~4 and 5 are obtained for on-resonance excitation with $\Delta=0$. One of our near-future goals is to measure absorption rise-time $\tau$ as a function of the optical depth as well as temperature for finite detuning of the excitation laser with  $\Delta \neq 0$.


\newpage




\begin{references}

\bibitem{dicke} R. H. Dicke, Coherence in Spontaneous Radiation Processes, Phys. Rev. {\bf 93}, 99 (1954).

\bibitem{haroche} M. Gross and S. Haroche, Superradiance: An Essay on the Theory of Collective Spontaneous Emission, Phys. Rep. {\bf 93}, 301 (1982).

\bibitem{scully1} M. O. Scully and A. A. Svdzinsky, The Super of Superradiance, Science {\bf 325}, 1510 (2009).

\bibitem{yelin} H. Ma, O. Rubies-Bigorda, and S. F. Yelin, Superradiance and Subradiance in a Gas of Two-level Atoms, arXiv:2205.15255 [quant-ph] (2022). 

\bibitem{francis} F. Robicheaux and D. A. Suresh, Beyond lowest order mean-field theory for light interacting with atom arrays, Phys. Rev. A {\bf 104}, 023702 (2021). 

\bibitem{adams} R. J. Bettles, S. A. Gardiner, and C. S. Adams, Cooperative Eigenmodes and Scattering in One-Dimensional Atomic Arrays, Phys. Rev. A {\bf 94}, 043844 (2016). 

\bibitem{jenkins} S. D. Jenkins and J. Ruostekoski, Controlled Manipulation of Light by Cooperative Response of Atoms in an Optical Lattice, Phys. Rev. A {\bf 86}, 031602 (2012).

\bibitem {ritsch} H. Zoubi and H. Ritsch, {\it  Metastability and Directional Emission Characteristics of Excitons in 1D Optical Lattices}, Europhys. Lett. {\bf 90}, 23001 (2010).

\bibitem{bachelard} C. E. Máximo,  R. Bachelard, F. E. A. dos Santos, and C. J. Villas-Boas, {\it Cooperative Spontaneous Emission via Renormalization Approach: Classical Versus Semi-Classical Effects},  arXiv:1906.05719 (2020). 

\bibitem{petrov} D. F. Kornovan, A. S. Sheremet, and M. I. Petrov, {\it Collective Polaritonic Modes in an Array of Two-Level Quantum Emitters Coupled to an Optical Nanofiber}, Phys. Rev. B {\bf 94}, 245416 (2016).

\bibitem{zanthier} D. Bhatti, R. Schneider, S. Oppel and J. von Zanthier , Directional Dicke Subradiance with Nonclassical and Classical Light Sources, Phys. Rev. Lett. {\bf 120}, 1136 (2018).

\bibitem{agarwal} R. Wiegner, J. von Zanthier, and G. S. Agarwal, Quantum-interference-initiated Superradiant and Subradiant Emission from Entangled Atoms, Phys. Rev. A {\bf 84}, 023805 (2011).

\bibitem{reitz} M. Reitz, C. Sommer, and C. Genes, Cooperative Quantum Phenomena in Light-Matter Platforms, PRX Quantum {\bf 3}, 010201 (2022).

\bibitem{bennett} A. Burgess, M. C. Waller, E. M. Gauger, and Robert Bennett, Phys. Rev. Lett. {\bf 134}, 113602 (2025). 

\bibitem{cirac} B. Windt, M. Bello, D. Malz, and J. Ignacio-Cirac, Effects of Retardation on Many-Body Superradiance in Chiral Waveguide QED, Phys. Rev. Lett. {\bf 134}, 173601 (2025). 

\bibitem{pohl} J. Kumlin, A. Srivastava, and Thomas Pohl, Superradiance of Strongly Interacting Dipolar Excitons in Moire Quantum Materials, Phys. Rev. Lett. {\bf 134}, 126901 (2025). 


\bibitem{feld}  N. Skribanowitz, I. P. Herman, J. C. MacGillivray, and M. S. Feld, Observation of Dicke Superradiance in Optically Pumped HF Gas, Phys. Rev. Lett. {\bf 30}, 309 (1973).

\bibitem{manassah} R. Friedberg, S. R. Hartmann, and J. T. Manassah, Frequency Shifts in Emission and Absorption by Resonant Systems of Two-level Atoms, Phys. Rep. {\bf 7}, 101 (1973).

\bibitem{bloch} J. Rui, D. Wei, A. Rubio-Abadal, S. Hollerith, J. Zeiher, Dan M. Stamper-Kurn, C. Gross, and I. Bloch, A Subradiant Optical Mirror Formed by a Single Structured Atomic Layer, Nature {\bf 583}, 369 (2020).

\bibitem{an} J. Kim, D. Yang, S. Oh, and K. An, Coherent Single-Atom Superradiance, Science 359, 662 (2018). 

\bibitem{gauthier} J. A. Greenberg and D. J. Gauthier, Steady-state, Cavityless, Multimode Superradiance in a Cold Vapor, Phys. Rev. A {\bf 86}, 013823 (2012). 

\bibitem{kuga} Y. Yoshikawa, Y. Torii, and T. Kuga, Superradiant Light Scattering from Thermal Atomic Vapors, Phys. Rev. Lett. {\bf 94}, 083602 (2005). 

\bibitem{ions} R. G. DeVoe and R. G. Brewer, Observation of Superradiant and Subradiant Spontaneous Emission of Two Trapped Ions, Phys. Rev. Lett. {\bf 76}, 2049 (1996).

\bibitem{molecules} B. McGuyer, M. McDonald, G. Iwata et al., Precise Study of Asymptotic Physics with Subradiant Ultracold Molecules, Nature Phys. {\bf 11}, 32 (2015).

\bibitem{diamond1} C. Bradac, M. T. Johnsson, M. V. Breugel, et al., Room-temperature Spontaneous Superradiance From Single Diamond Nanocrystals, Nat. Commun. {\bf 8}, 1205 (2017).

\bibitem{diamond2} A. Angerer, K. Streltsov, T. Astner, T. et al., Superradiant Emission From Color Centers in Diamond, Nat. Phys. {\bf 14}, 1168 (2018).

\bibitem{superconducting} Z. Wang et al., Controllable Switching Between Superradiant and Subradiant States in a 10-qubit Superconducting Circuit, Phys. Rev. Lett. {\bf 124}, 013601 (2020).

\bibitem{quantumdot} C. Zhu, S. C. Boehme, L. G. Feld, A. Moskalenko, D. N. Dirin, R. F. Mahrt, T. Stoferle, M. I. Bodnarchuk, A. L. Efros, P. C. Sercel, M. V. Kovalenko, and G. Raino, Single-photon Superradiance in Individual Caesium Lead Halide Quantum Dots, Nature {\bf 626}, 535 (2024). 

\bibitem{kaiser1} W. Guerin, M. O. Araujo, and R. Kaiser, Subradiance in a Large Cloud of Cold Atoms, Phys. Rev. Lett. 116, 083601 (2016).

\bibitem{kaiser2} P. Weiss, M. O Araújo, R. Kaiser and W. Guerin, Subradiance and Radiation Trapping in Cold Atoms, New J. Phys. {\bf 20}, 063024 (2018).

\bibitem{kaiser3} T. Bienaime, N. Piovella, and R. Kaiser, Controlled Dicke Subradiance from a Large Cloud of Two-Level Systems, Phys. Rev. Lett. {\bf 108}, 123602 (2012). 

\bibitem{kaiser4} A. Cipris, N. A. Moreira, T. S. E. Santo, P. Weiss, C. J. Villas-Boas, R. Kaiser, W. Guerin, and R. Bachelard, Subradiance with Saturated Atoms: Population Enhancement of the Long-Lived States, Phys. Rev. Lett. {\bf 126}, 103604 (2021). 

\bibitem{kaiser5} P. Weiss, A. Cipris, M. O. Araujo, R. Kaiser, and W. Guerin, Robustness of Dicke subradiance against thermal decoherence, Phys. Rev. A {\bf 100}, 033833 (2019). 

\bibitem{browaeys1} Giovanni Ferioli, Antoine Glicenstein, Loic Henriet, Igor Ferrier-Barbut , and Antoine Browaeys, Storage and Release of Subradiant Excitations in a Dense Atomic Cloud, Phys. Rev. X {\bf 11}, 021031 (2021). 

\bibitem{browaeys2} A. Glicenstein, G. Ferioli, A. Browaeys, and I Ferrier-Barbut, From superradiance to subradiance: exploring the many-body Dicke ladder, Opt. Lett. {\bf 47}, 1541 (2022). 

\bibitem{browaeys3} G. Ferioli, A. Glicenstein, F. Robicheaux, R. T. Sutherland, A. Browaeys, and I. Ferrier-Barbut, Laser-Driven Superradiant Ensembles of Two-Level Atoms near Dicke Regime, Phys. Rev. Lett. {\bf 127}, 243602 (2021). 

\bibitem{superabsorption} D. Yang, S. Oh, J. Han, G. Son, J. Kim, J. Kim, M. Lee, and  K. An, Realization of Superabsorption by Time Reversal of Superradiance, Nature Photonics {\bf 15}, 272 (2021). 


\bibitem{dipto} D. Das, B. Lemberger, and D. D. Yavuz, Subradiance and Superradiance-to-Subradiance Transition in Dilute Atomic Clouds, Phys. Rev. A {\bf 102}, 043708 (2020).

\bibitem{davidexp} D. C. Gold, P. Huft, C. Young, A. Safari, T. G. Walker, M. Saffman, and D. D. Yavuz, Spatial Coherence of Light in Collective Spontaneous Emission, PRX Quantum {\bf 3}, 010338 (2022). 

\bibitem{yavuz_coherence} D. D. Yavuz, A. Yadav, D. C. Gold, T. G. Walker, and M. Saffman, Numerical Study of the Spatial Coherence of Light in Collective Spontaneous Emission, Phys. Rev. A {\bf 110}, 043705 (2024). 

\bibitem{rudhy} A. Yadav and D. D. Yavuz, Analytical and Numerical Studies
of Subradiance-only Collective Decay from Dilute Atomic Ensembles, Phys. Rev. A {\bf 110}, 023709 (2024).

\bibitem{david_inprep} D. C. Gold, S. Carpenter, T. G. Walker, M. Saffman, and D. D. Yavuz, Subradiance and Subradiance-to-Radiation-Trapping Transition in Dilute Atomic Ensembles (in preparation). 

\bibitem{yelin1} E. Shahmoon, D. S. Wild, M. D. Lukin, and S. F. Yelin, Cooperative Resonances in Light Scattering from Two-Dimensional Atomic Arrays, Phys. Rev. Lett. {\bf 118}, 113601 (2017).

\bibitem{yelin2} O. Rubies-Bigorda, S. Ostermann, and S. F. Yelin, Dynamic population of multiexcitation subradiant states in incoherently excited atomic arrays, Phys. Rev. A {\bf 107}, L051701 (2023).

\bibitem{francis1} R. T. Sutherland and F. Robicheaux, Coherent Forward Broadening in Cold Atom Clouds, Phys. Rev. A {\bf 93}, 023407 (2016).  

\bibitem{francis2} D. A. Suresh and F. Robicheaux, Photon-induced atom recoil in collectively interacting planar arrays, Phys. Rev. A {\bf 103}, 043722 (2021).

\bibitem{francis3} F. Robicheaux, Theoretical study of early-time superradiance for atom clouds and arrays, Phys. Rev. A {\bf 104}, 063706 (2021). 

\bibitem{ritsch2} H. Zoubi and H. Ritsch, {\it Lifetime and Emission Characteristics of Collective Electronic Excitations in Two-Dimensional Optical Lattices}, Phys. Rev. A {\bf 83}, 063831 (2011).

\bibitem{ballantine} K. E. Ballantine and J. Ruostekoski, Quantum Single Photon Control, Storage, and Entanglement Generation with Planar Atomic Arrays, PRX Quantum {\bf 2}, 040362
(2021).

\bibitem{garcia} E. Sierra, S. J. Masson, and A. Ansejo-Garcia, Dicke Superradiance in Ordered Lattices: Dimensionality Matters, Phys. Rev. Research {\bf 4}, 023207 (2022). 


\bibitem{agarwal2} D. Bhatti, R. Schneider, S. Oppel and J. von Zanthier , Directional Dicke Subradiance with Nonclassical and Classical Light Sources, Phys. Rev. Lett. {\bf 120}, 1136 (2018).

\bibitem{agarwal3} J. Xu, S. Chang, Y. Tang, S. Zhu, G. S. Agarwal, Hyperradiance accompanied by nonclassicality, Phys. Rev. A {\bf 96}, 013839 (2017). 

\bibitem{agarwal4} M. Pleinert, J. von Zanthier, and G. S. Agarwal, Hyperradiance from collective behavior of coherently driven atoms, Optica {\bf 4}, 779 (2017). 

\bibitem{masson} S. J. Masson, J. P. Covey, S. Will, and A. Asenjo-Garcia, PRX Quantum {\bf 5}, 010344 (2024). 

\bibitem{kimble} A. Asenjo-Garcia, M. Moreno-Cardoner, A. Albrecht, H.J. Kimble, and D. E. Chang, Exponential Improvement in Photon Storage Fidelities Using Subradiance and “Selective Radiance” in Atomic Arrays, Phys. Rev. X {\bf 7}, 031024 (2017).

\bibitem{browaeys4} G. Ferioli, A. Glicenstein, I. Ferrier-Barbut and
A. Browaeys, A non-equilibrium superradiant phase
transition in free space, Nature Physics {\bf 19}, 1345 (2023).

\bibitem{yan} Z. Yan, J. Ho,  Y. Lu, S. J. Masson,
A. Asenjo-Garcia, and D. M. Stamper-Kurn, Superradiant and Subradiant Cavity Scattering by Atom Arrays, Phys. Rev. Lett. {\bf 131}, 253603 (2023). 

\bibitem{hung} X. Zhou, H. Tamura, T. H. Chang, and C. L. Hung, Trapped Atoms and Superradiance on an Integrated Nanophotonic Microring Circuit, Phys. Rev. X {\bf 14}, 031004 (2024). 

\bibitem{liedl} C. Liedl, F. Tebbenjohanns, C. Bach, S. Pucher, A. Rauschenbeutel, and P. Schneeweiss, Observation of Superradiant Bursts in a Cascaded Quantum System, Phys. Rev. X {\bf 14}, 011020 (2024). 

\bibitem{ben} B. Lemberger and D. D. Yavuz, Effect of Correlated Decay on Fault-tolerant Quantum Computation, Phys. Rev. A {\bf 96}, 062337 (2017).

\bibitem{bellando} L. Bellando, A. Gero, E. Akkermans, and R. Kaiser, Roles of Cooperative Effects and Disorder in Photon Localization: The Case of a Vector Radiation Field, Eur. Phys. J. B 94:49 (2021). 

\bibitem{light_anderson} S. E. Skipetrov and I. M. Sokolov, Absence of Anderson Localization of Light in a Random Ensemble of Point Scatterers, Phys. Rev. Lett. {\bf 112}, 023905 (2014).

\bibitem{eberly} L. Allen and J. H. Eberly, Optical Resonance and Two-level Atoms, Dover Publications (1987). 

\bibitem{adams_dipole} L. Weller, R. J. Bettles, P. Siddons, C. S. Adams, and I. G. Hughes, Absolute Absorption on Rubidium D1 Line: Including Resonant Dipole-Dipole Interactions, J. of Phys. B: At. Mol. Opt. Phys. {\bf 44}, 195006 (2011). 

\bibitem{rosa} J. G. Rosa, Superradiance in the Sky, Phys. Rev. D {\bf 95}, 064017 (2017). 

\bibitem{witek} H. Witek, V. Cardoso, A. Ishibashi, and Ulrich Sperhake, Superradiant Instabilities in Astrophysical Systems, Phys. Rev. D {\bf 87}, 043513 (2013).

\bibitem{sarmah} P. Sarmah, H. Verma, K. Cheung, and J. Silk, Effects of Superradiance in Active Galactic Nuclei, arXiv:2404.09955 [astro-ph.HE] (2024). 

\bibitem{teo} M. Baryakhtar, R. Lasenby, and M. Teo, Black Hole Superradiance Signatures of Ultralight Vectors, Phys. Rev. D {\bf 96}, 035019 (2017).

\bibitem{zhou} F. M. Chang, H. Gao, V. Jaramillo, X. Meng, and S. Y. Zhou, Boson Star Superradiance with Spinning Effects and in Time Domain, Phys. Rev. D. {\bf 111}, 044053 (2025).


\bibitem{mandel_wolf} L. Mandel and E. Wolf, Optical Coherence and Quantum Optics (Cambridge University Press, 1995).

\bibitem{mark_review} M. Saffman, Quantum Computing with Neutral Atoms, National Science Review {\bf 6}, 24 (2019). 

\bibitem{browaeys_large} G. Pickard {\it{et al.}}, Rearrangement of Individual Atoms in a 2000-Site Optical-Tweezer Array at
Cryogenic Temperatures, Phys. Rev. Applied {\bf 22}, 024073 (2024). 

\bibitem{lukin_logical} D. Bluvstein {\it{et al.}}, Logical Quantum Processor Based on Reconfigurable Atom Arrays, Nature {\bf 626}, 58 (2024).  



\end{references}
\end{document}